\documentclass[10pt,journal,compsoc]{IEEEtran}
\newif\ifpeerreview

\peerreviewfalse

\usepackage[nocompress]{cite}
\usepackage{url}
\usepackage{amsmath,amssymb,graphicx}

\usepackage{lipsum} 

\usepackage{multirow}
\usepackage{booktabs}
\usepackage{url}
\usepackage[pagebackref=true,breaklinks=true,colorlinks,bookmarks=false,urlcolor=blue,citecolor=blue,linkcolor=blue]{hyperref}

\usepackage[table]{xcolor} 
\definecolor{Gray}{gray}{0.9} 
\definecolor{rblue}{rgb}{0,0.5,1}

\newcommand{\ourframework}{FoundCAC}
\usepackage{pifont} 

\usepackage{caption}
\usepackage{wrapfig}
\usepackage{enumitem}
\usepackage{paralist}

\usepackage{ragged2e}

\usepackage[switch]{lineno}

\usepackage{icomma}

\usepackage{marvosym}

\title{Towards Blind Lens Aberration Correction via Large LensLib Pre-training and Discrete Degradation Priors}
\author{Xiaolong~Qian$^{*}$,
Qi~Jiang$^{*}$,
Yao~Gao,
Lei~Sun$\textsuperscript{\Letter}$,
Kailun~Yang,
Xian~Wang,
Zhonghua~Yi,
Wenyong~Li,
Ming-Hsuan~Yang,
Luc~Van~Gool,
and~Kaiwei~Wang$\textsuperscript{\Letter}$
\IEEEcompsocitemizethanks{
\IEEEcompsocthanksitem X. Qian, Q. Jiang, Y. Gao, L. Sun, X. Wang, Z. Yi, W. Li, and K. Wang are with the National Research Center for Optical Instrumentation, Zhejiang University, Hangzhou 310027, China.
\IEEEcompsocthanksitem L. Sun is also with INSAIT, Sofia University ``St. Kliment Ohridski'', Sofia 1784, Bulgaria.
\IEEEcompsocthanksitem K. Yang is with the School of Artificial Intelligence and Robotics, Hunan University, Changsha 410012, China.
\IEEEcompsocthanksitem K. Yang is also with the National Engineering Research Center of Robot Visual Perception and Control Technology, Hunan University, Changsha 410082, China.
\IEEEcompsocthanksitem $^{*}$Equal contribution.
\IEEEcompsocthanksitem $\textsuperscript{\Letter}$Correspondence (E-mail: \href{leo_sun@zju.edu.cn}{leo\_sun@zju.edu.cn}, \href{wangkaiwei@zju.edu.cn}{wangkaiwei@zju.edu.cn}).
}
}

\begin{document}

\IEEEtitleabstractindextext{%
\begin{abstract} \justifying
Emerging deep-learning-based lens library pre‑training (LensLib-PT) pipeline offers a new avenue for blind lens aberration correction by training a universal neural network, demonstrating strong capability in handling diverse unknown optical degradations. This work proposes FoundCAC, a universal foundational framework that resolves two challenges hindering the generalization of existing pipelines: the difficulty of scaling training data and the absence of prior guidance characterizing optical degradation. To improve data scalability, we expand the design specifications to increase degradation diversity and construct AODLibpro, a large-scale lens library using stratified sampling over spatial-variation patterns and degradation severity. In terms of model design, to leverage Point Spread Functions (PSFs) as guidance while maintaining the blind paradigm, we propose a multi-stage vector-quantized representation learning scheme. This paradigm is specifically designed to construct a Latent PSF Representation (LPR), explicitly encoding complex continuous PSFs into a discrete degradation prior to regularize the highly ill-posed restoration process. Through a simple yet effective codebook-freezing strategy, our framework leverages the discrete prior to elevate full-shot restoration performance and unlock highly efficient few-shot adaptation for unseen lenses. Experiments on synthetic LensLib, real-design simulations, and real-captured lenses show that our framework achieves state-of-the-art zero-shot performance under complementary evaluation protocols, while enabling highly efficient few-shot adaptation for specific lenses. The source code and datasets will be made publicly available at \href{https://github.com/zju-jiangqi/FoundCAC}{FoundCAC}.
\end{abstract}

}

\maketitle

\begin{figure*}[!h]
  \centering
  \includegraphics[width=1.0\linewidth]{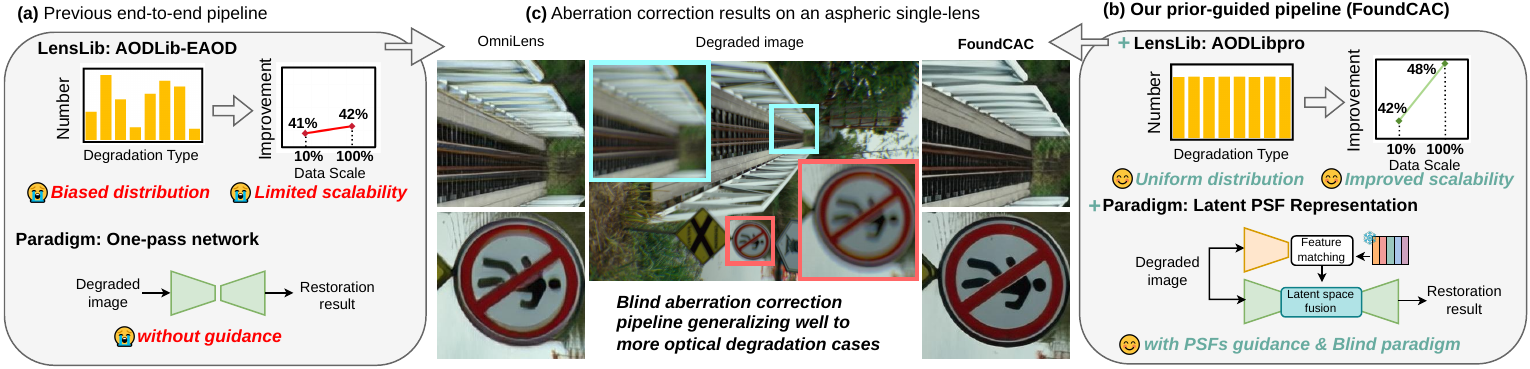}
  \vspace{-2.0em}
  \caption{This work addresses the challenges of the current LensLib-PT pipeline. 
  (a) Previous end-to-end pipelines (\textit{e.g.}, OmniLens~\cite{jiang2024flexible}) often suffer from limited data scalability due to biased training distributions and the absence of explicit physical prior guidance.
  (b) In FoundCAC, the proposed AODLibpro reveals a uniform distribution contributing to improved scalability, while the latent PSF representation provides effective prior guidance with a blind paradigm. 
  (c) FoundCAC effectively restores severe aberrations where conventional paradigms struggle.}
  \label{fig:teaser}
  \vspace{-1.0em}
\end{figure*}

\IEEEraisesectionheading{
  \section{Introduction}
  \label{sec:introduction}
}
\IEEEPARstart{L}{ens} aberrations, typically arising from compromised image quality optimization due to design trade-offs for specific requirements, \textit{e.g.}, minimalist optical systems~\cite{heide2013high, peng2019learned}, or lenses on mobile devices~\cite{chen_mobile_2023}, and manufacturing/assembly errors~\cite{liu2024application} in complex systems, introduce blur to the captured images.
This blur is also referred to as optical degradation~\cite{chen2021optical}, characterized by its distinctive spatially-varying nature where degradation varies across Field-of-Views (FoVs) and exhibits diverse patterns depending on optical path, representing a fundamental image quality issue but has received limited attention in the learning and vision literature. 
With the advancement of image processing, computational post-processing~\cite{schuler2011non} has become a mainstream pipeline, also known as computational aberration correction.
Unlike non-blind methods that rely on precise Point Spread Functions (PSFs) calibration~\cite{chen2025physics}, the blind pipeline~\cite{schuler2012blind} offers more flexible and user-friendly advantages for users without optical expertise, where only the captured images are required for high-quality results. 
Recently, the deep learning-based Lens Library Pre-Training pipeline (LensLib-PT) has emerged as a powerful blind aberration correction paradigm~\cite{gong2024physics}.
A universal network is trained to learn the mapping from diverse aberration distributions to clear images, demonstrating advantages over traditional blind deconvolution methods~\cite{eboli2022fast} in terms of generalization to different optical degradation types.
Representative frameworks, such as OmniLens~\cite{jiang2024flexible}, have further pushed this boundary by leveraging large-scale libraries of automatically designed lens samples.

However, despite these advances, the broader application of the general LensLib-PT paradigm remains constrained by two fundamental bottlenecks: 1) the inherent bias and limited \textit{scalability} of current synthetic lens libraries; and 2) the absence of explicit physical \textit{guidance} in existing end-to-end restoration models to adaptively handle highly diverse aberration distributions.
To address these dual bottlenecks, we propose the Foundational Computational Aberration Correction (FoundCAC) framework, featuring joint innovations in data construction and model architecture.

Regarding the data bottleneck, constructing a truly unbiased and comprehensive lens library remains challenging. Existing large-scale datasets, such as AODLib-EAOD~\cite{jiang2024flexible}, still exhibit a biased optical degradation distribution. 
This bias primarily stems from their reliance on incomplete design specifications and overly simplified sampling bases (\textit{e.g.}, average RMS spot radius), which fail to accurately capture the complex spatial-variation patterns of real-world aberrations~\cite{zhou2024revealing,qian2026unicac}. 
Consequently, simply increasing the scale of such libraries yields diminishing returns due to the structural absence of specific degradation types.
To resolve this at the data level, FoundCAC introduces a novel hybrid sampling basis to cover diverse optical degradation patterns. 
Instead of relying on lens design indicators, this basis directly quantifies the severity and spatial variation trends from per-FoV imaging results.
As depicted at the top of Figure~\ref{fig:teaser}~(b), supplemented with enriched design specifications, the hybrid sampling contributes to the \textit{large LensLib AODLibpro} with uniform aberration distributions without bias.
The trained universal model can benefit from its scalability and achieve significant improvements with large‑scale data.
In addition, a synthetic benchmark is also established on the sampled lenses for evaluating aberration correction methods in terms of model design, which is the first benchmark in this field for comprehensive evaluation across diverse aberration patterns.

Besides constructing scalable data resources, an equally important challenge lies in designing models that can effectively handle diverse degradation types.
Introducing degradation prior representations as guidance is a crucial design for such universal models~\cite{potlapalli2023promptir,huuniversal}. 
In aberration correction, PSFs serve as the fundamental physical representation of optical degradation.
Whether through explicit deconvolution~\cite{lin2022non} or implicit representation embedding~\cite{jiang2024minimalist}, correction results can be effectively improved. 
However, these approaches require precise PSFs of the target lenses, which is infeasible in a blind pipeline.
To bridge this gap, FoundCAC introduces a novel multi-stage training paradigm specifically designed to construct a Latent PSF Representation (LPR).
In the first stage, we explicitly encode the diverse, continuous PSFs from AODLibpro into a discrete Vector-Quantized (VQ) codebook, establishing the LPR as a comprehensive physical dictionary of optical degradations. 
In the second stage, the blind correction network learns to predict and retrieve these discrete latent priors directly from degraded images, providing robust structural constraints for the restoration process without requiring actual GT PSFs.

Beyond zero-shot generalization, the discrete LPR enables robust target adaptation via a universal codebook-freezing strategy. By strictly anchoring the network to the established optical prior, this mechanism yields dual benefits: in \textit{full-shot} scenarios, it provides a strong initialization that effectively boosts the final restoration quality; under \textit{few-shot} conditions, it acts as an explicit structural regularizer that prevents overfitting, ensuring highly efficient adaptation.

Extensive experiments across diverse types of minimalist optical systems, misaligned lenses, high-end lenses, and our benchmark demonstrate that \ourframework~achieves state-of-the-art generalization capacity in \textit{zero-shot} blind correction.
Taking an aspheric single-lens as an example, Figure~\ref{fig:teaser}~(c) illustrates that \ourframework~effectively restores severe aberrations where conventional paradigms struggle. 
The main contributions of this work are:
\begin{compactitem}
    \item We propose \ourframework, a LensLib-PT framework for blind aberration correction. It introduces a multi-stage Vector-Quantized (VQ) representation learning paradigm to construct a Latent PSF Representation (LPR), which provides discrete priors and bypasses the need for precise target PSFs.
    \item We construct AODLibpro, a scalable, de-biased lens library driven by a novel hybrid sampling strategy, and establish the first comprehensive synthetic benchmark for diverse aberration patterns.
    \item Extensive experiments show that \ourframework~achieves state-of-the-art zero-shot performance under complementary evaluation on synthetic LensLib, real-design simulations, and real-captured images. With a simple codebook-freezing strategy, it also provides an effective initialization for \textit{full-shot} target adaptation and improves \textit{few-shot} adaptation efficiency for specific lenses.
\end{compactitem}

\section{Related Work}
\label{sec:Related Work}
\noindent\textbf{Lens aberration correction} through post-processing is widely applied in computational imaging, commonly used for Minimalist Optical System (MOS) imaging~\cite{wei2024computational,qian2025towards,tseng2021neural,qian2025deveiler} and image quality enhancement for mobile devices~\cite{chen_mobile_2023,schuler2012blind}.
The non-blind methods with lens-specific paradigm~\cite{chen2021optical,yanny2022deep} represent the current mainstream, but the repeated complex calibration~\cite{chen2025physics} and model training~\cite{chen2021extreme_quality} for each different lens make it unfriendly to users without optical backgrounds. 
Blind pipelines offer a flexible alternative, requiring only degraded images.
Traditional methods rely on kernel estimation combined with natural image priors for deconvolution~\cite{2011Modeling,schuler2012blind,yue2015blind,eboli2022fast}, but are restricted to mild aberrations and struggle to generalize across diverse lenses. 
The recent data-driven Lens Library Pre-Training (LensLib-PT) pipeline addresses this issue by training a universal model on a LensLib covering diverse aberrations. 
Early research works~\cite{li2021universal,gong2024physics} leverage a few manually collected lenses, where the scalability constraints result in limited coverage. 
While Zernike-based databases~\cite{hu2021image,jiang2024computational} contribute to data expansion, they suffer from shortcomings in the realism of distributions. 
In comparison, AODLib-EAOD constructed by EAOD algorithms in OmniLens achieves a balance between data scale and aberration distribution authenticity~\cite{jiang2024flexible}. 
Nevertheless, expanding the scale of AODLib-EAOD yields limited improvements, which is due to the data bias introduced by limited specifications and the simple sampling basis.
To this end, this work incorporates more design specifications into AOD for broader coverage.
A hybrid sampling basis is then designed based on the quantification of both severity and spatial-varying trends of optical degradation.

{\noindent\textbf{Discrete degradation priors and adaptation}} is widely studied in All-in-One Image Restoration (AIO-IR) to guide models in processing different degradation types~\cite{jiang2025survey}, yet remains underexplored in lens aberration correction.
AIO-IR methods typically employ visual prompts~\cite{potlapalli2023promptir,ma2023prores}, contrastive learning~\cite{li2022all}, text semantic prompts~\cite{ai2024multimodal}, large model based feature extraction~\cite{zhang2025perceive}, and pretext tasks~\cite{huuniversal} to characterize categorical information of different degradations. 
Unlike AIO-IR, different degradations in lens aberration correction all stem from convolution-induced image blur, varying primarily in severity and spatial-varying patterns, making category-based designs inapplicable.
Given that PSFs directly characterize such degradations, several works propose using PSF information to guide aberration correction~\cite{li2021universal,lin2022non,jiang2024minimalist,luo2024correcting}. 
However, requiring precise PSFs during inference prevents these efforts from realizing blind operation.
To leverage PSF-based guidance while achieving blind operation, this work explores predicting PSF-related information from degraded images. 
In particular, VQVAE~\cite{van2017neural} provides a powerful framework by utilizing VQ codebooks~\cite{esser2021taming} in latent space to store discrete key features, a strategy that has been successfully investigated to represent optical degradation priors for guiding aberration correction~\cite{chen_mobile_2023,jiang2025representing}.
Inspired by this, we propose a multi-stage foundational paradigm leveraging vector-quantized codebooks to construct a latent PSF representation. 
Unlike previous implicit representations, our approach constructs an explicit physical dictionary of optical degradations through joint optimization. 
This strategy not only enables accurate blind prior retrieval for superior zero-shot performance but also provides a robust initialization for highly efficient target-specific adaptation on unseen lenses.

\section{Proposed Method}
\label{sec:Proposed Method}
\subsection{Motivation}
\label{sec:motivation}
Building a LensLib-PT-based blind aberration correction framework hinges on two axes: 1) the coverage and aberration distribution of LensLib determine the application scope of the trained model; and 2) the model paradigm determines whether it can leverage the potential of the data. Therefore, we consider both aspects to train a foundational model with stronger generalization ability.

\noindent\textbf{Data foundation construction.} Conventional LensLib-PT pipelines often struggle with unseen lens configurations (\textit{e.g.}, the aspheric lens shown in Figure~\ref{fig:teaser}~(c)), indicating that incomplete design specifications during the lens source generation stage lead to insufficient coverage of complex optical degradation distributions. 
Therefore, a broader range of specifications must be incorporated to fundamentally increase the diversity of aberration coverage. 
Furthermore, as illustrated in Figure~\ref{fig:teaser}~(a), directly expanding the scale of existing datasets (\textit{e.g.}, AODLib-EAOD) yields marginal improvements, highlighting a severe scalability bottleneck. 
This limitation primarily stems from their reliance on an RMS-based sampling basis. 
Since RMS-based metrics can neither accurately describe the spatial-varying patterns of optical degradation nor reflect its severity~\cite{zhou2024revealing}, simply increasing the data scale fails to compensate for the structural absence of certain degradation types. Consequently, designing a novel hybrid sampling basis that jointly quantifies both the spatial-varying properties and the severity of optical degradations becomes imperative.

\begin{table}[!t]
    \begin{center}
        \caption{Results of existing model paradigms on the benchmark set up in Section~\ref{sec:lib}. Settings are detailed in the appendix.}
        \vspace{-1em}
        \label{tab:motivation}
        \resizebox{1.0\linewidth}{!}{
        \renewcommand{\arraystretch}{1.2}
        \setlength{\tabcolsep}{2.5mm}{ 
            \begin{tabular}{lcccc}
            \toprule[0.17em]
            \textbf{Paradigm} & \textbf{Blind} & \textbf{PSNR $\uparrow$} & \textbf{SSIM $\uparrow$} & \textbf{LPIPS $\downarrow$} \\ 
            \midrule[0.17em]
            Baseline & \checkmark & 28.46 & 0.873 & 0.1318 \\
            GT-PSFs-guided & \ding{55} & 28.73 & 0.873 & 0.1269 \\
            PSFs prediction & \checkmark & 28.25 & 0.861 & 0.1283 \\
            PSFs feature prediction & \checkmark & 28.42 & 0.867 & 0.1287 \\
            \bottomrule[0.17em]
            \end{tabular}
        }
        }
    \end{center}
    \vspace{-2em}
\end{table}

\noindent\textbf{Model paradigm evolution.} Table~\ref{tab:motivation} shows the performance of several existing aberration correction model paradigms, among which using Ground-Truth (GT) PSFs to guide the model demonstrates great potential. 
However, for the blind paradigm, such precise PSF information is unavailable. 
Directly predicting PSFs from degraded images (the third row of Table~\ref{tab:motivation}) is an intuitive pipeline, but such a task is revealed to be a highly ill-posed inverse problem~\cite{joshi2008psf,rego2021robust}. 
Even predicting continuous PSF features in the latent space (the fourth row of Table~\ref{tab:motivation}) alleviates this difficulty but still lacks explicit structural constraints. Since the continuous latent space is essentially unbounded, it remains challenging for the network to reliably predict valid degradation representations, especially when facing diverse unseen lenses.
To this end, instead of relying on unconstrained continuous feature prediction, we formulate the extraction of degradation priors as a \textit{discrete retrieval} process. 
Specifically, we construct a Latent PSF Representation (LPR) as a discrete codebook that explicitly encodes degradation priors.
By replacing unconstrained continuous regression with a more tractable discrete retrieval process, this quantization significantly reduces the ambiguity inherent in highly ill-posed blind correction.
Consequently, the proposed LPR provides physics-informed guidance for zero-shot generalization, while naturally enabling a unified adaptation strategy that boosts full-shot restoration performance and ensures highly efficient few-shot fine-tuning for unseen target lenses.

\begin{figure*}[!h]
  \centering
  \includegraphics[width=1.0\textwidth]{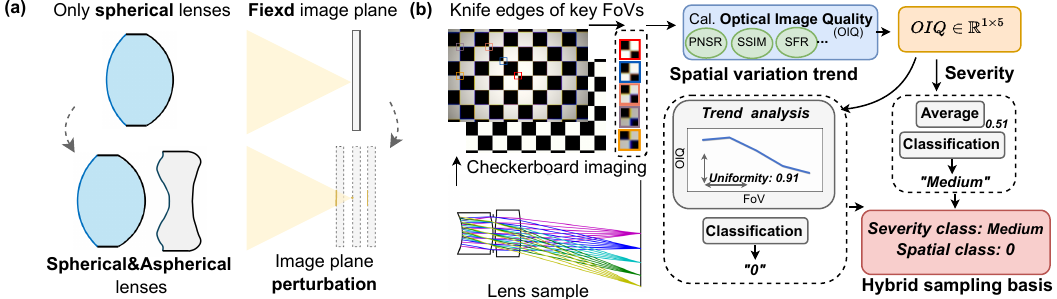}
  \vspace{-2em}
  \caption{Illustration of the key designs in constructing AODLibpro. We expand surface type and imaging distance specifications in (a) to realize a broader set of optical degradation patterns during lens source generation; and quantify degradation severity and spatial variation trends via image quality assessment in (b), yielding a hybrid sampling that covers plausible optical degradation patterns.}
  \label{fig:aodlibpro}
  \vspace{-1em}
\end{figure*}

\subsection{Data Foundation: Construction of AODLibpro}
\label{sec:lib}
\noindent\textbf{Generation of lens source.} To preserve diversity and realism in LensLib, we retain EAOD~\cite{jiang2024flexible} for generating the lens source under different design specifications.
We refer readers to~\cite{jiang2024flexible} for details on EAOD.
Building on this, we incorporate two previously omitted but important specifications to further diversify the lens source as shown in Figure~\ref{fig:aodlibpro} (a): (i) aspheric surface to broaden reachable degradation patterns, where high‑order aspheric coefficients are added to lens parameters and EAOD ray tracing is extended to handle aspheric surfaces following~\cite{qian2026unicac}; and (ii) image distance, which shifts the focal field and shapes spatial degradation characteristics. 
To model the latter, we perturb the image distance of EAOD-optimized lenses within the depth of field with probability $\gamma$ to generate additional variants.
Finally, all specifications are fed into EAOD to search for candidate solutions that form the lens source. 
More details on the two specifications can be found in the appendix.

\noindent\textbf{Hybrid sampling basis.}
Since traditional design indicators do not fully reflect image-domain optical degradation~(\S~\ref{sec:motivation}), we quantify it directly from lens imaging results.
We measure per-FoV degradation severity and its spatial variation, forming a hybrid sampling basis for balanced sampling across diverse degradation patterns. 
As shown in Figure~\ref{fig:aodlibpro}~(b), we use a degraded checkerboard image and its paired GT as the quantification basis.
Five knife-edge image patches sampled from the center to the periphery FoVs are cropped for quantification. 
Following the comprehensive benchmark~\cite{qian2026unicac}, we adopt the Optical Image Quality (OIQ) metric to measure per-FoV degradation severity, which is calculated as a weighted sum of fidelity metrics (PSNR and SSIM) and an optical metric (SFR).
The average OIQ across the $5$ FoVs is applied as the overall degradation severity of the target lens. 
For the entire lens source, we analyze the per‑sample average OIQ distribution and partition the severity into $3$ Severity-Classes as \textit{Strong}, \textit{Medium}, and \textit{Mild}.
Furthermore, we categorize the spatial variation patterns of optical degradation into 6 Spatial-Classes based on OIQ trends over FoVs, building upon the spatial uniformity evaluation introduced in~\cite{qian2026unicac}.
First, using the variance and mean of per‑FoV OIQ, we compute the coefficient of variation to measure spatial uniformity $U_{S}$.
Samples with values above a threshold $\alpha$ are labeled the {``spatial-uniform'' Spatial‑Class.} 
For $U_{S}<\alpha$, {$5$ additional Spatial-Classes} are defined by {the FoV of peak OIQ and the monotonicity of OIQ changes}. 
Detailed OIQ computation and Spatial-Class definitions are provided in the appendix.

\noindent\textbf{Construction of AODLibpro.}
Leveraging the proposed hybrid sampling basis, we construct AODLibpro by uniformly sampling the lens source to ensure balanced coverage across severity and spatial-variation patterns. 
Specifically, we form $18$ sub-classes by crossing the severity and spatial dimensions. 
From each sub-class, $m_{1}$ and $m_{2}$ instances are sampled to build the \texttt{train} and \texttt{test} sets, respectively. 
The former provides large-scale supervision, while the latter serves as a comprehensive benchmark for evaluating correction models across diverse aberration distributions.

\begin{figure*}[!t]
  \centering
  \includegraphics[width=1.0\textwidth]{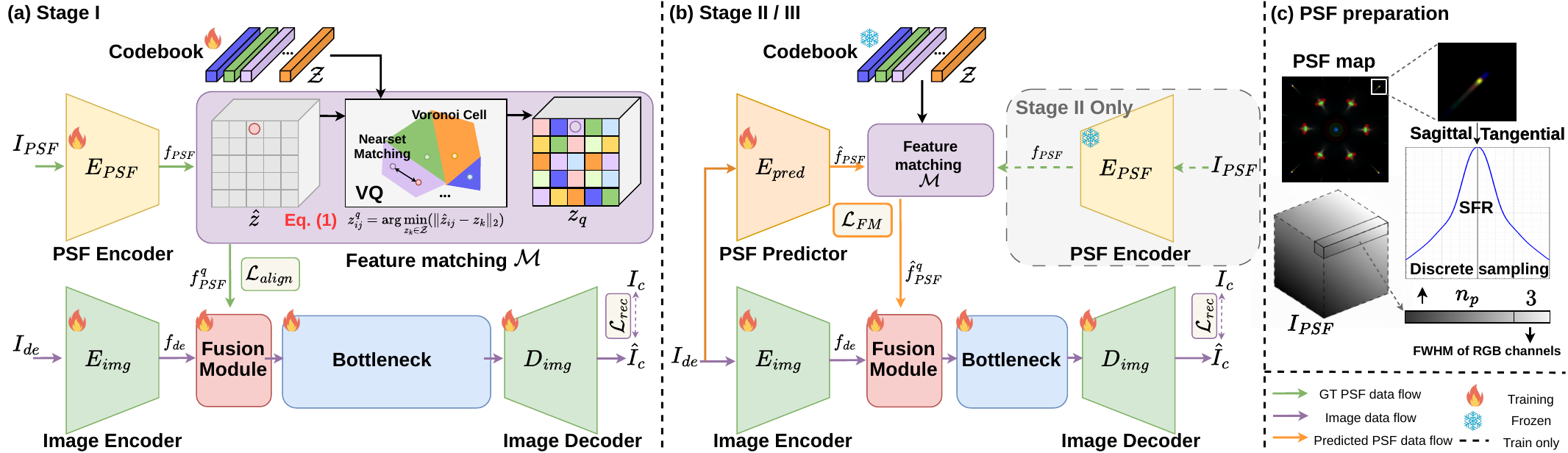}
  \vspace{-2em}
  \caption{Overview of the proposed~\ourframework~framework. 
  (a) In stage I, we construct the latent PSF representation via vector quantization, encoding continuous PSFs into a discrete physical degradation dictionary. 
  (b) Stage II trains a blind predictor to retrieve these discrete priors directly from degraded images for zero-shot restoration. Stage III performs target-specific adaptation by freezing the codebook, utilizing it as a structural regularizer for highly efficient fine-tuning. 
  (c) The preparation of ${I}_{PSF}$ serves as physical guidance for Stage I.}
  \label{fig:framework}
  \vspace{-1em}
\end{figure*}

\subsection{Stage I: GT-Guided Prior Pre-training}
\label{sec:stage1}
Unlike existing methods that rely on unconstrained continuous latent features, our framework formulates degradation prior extraction as a discrete retrieval task. 
In Stage I, we jointly train the restoration network and learn the proposed Latent PSF Representation (LPR) with a VQ codebook.

\noindent\textbf{Quantization of degradation priors.} 
As shown in Figure~\ref{fig:framework}, this stage takes the degraded image $I_{de}\in\mathbb{R}^{H{\times}W{\times}3}$ and its corresponding GT PSF map $I_{PSF}\in\mathbb{R}^{H{\times}W{\times}N_{p}}$ as inputs. 
The PSF map is processed by a PSF encoder $E_{PSF}$ to extract continuous spatial features $f_{psf}$. 
To align with the quantization formulation, we explicitly denote these continuous features as the latent representations $\hat{z} = f_{psf} \in\mathbb{R}^{h{\times}w{\times}n_{z}}$.
To establish the discrete LPR, we introduce a learnable codebook $\mathcal{Z}=\{z_k\}_{k=1}^{K}\in\mathbb{R}^{n_{z}}$. 
For each spatial element ${\hat{z}}_{ij}$, the quantization process identifies its nearest neighbor in $\mathcal{Z}$ to obtain the discrete representation $z^q\in\mathbb{R}^{h{\times}w{\times}n_{z}}$:
\begin{equation}
\label{eq:codematching}
z^{q}_{ij} = \arg\min\limits_{z_k\in\mathcal{Z}}(\Vert{\hat{z}_{ij}-z_k}\Vert_2).
\end{equation}

This operation explicitly maps the continuous, unbounded PSF features into a constrained discrete dictionary. 
This dictionary constitutes the proposed LPR, providing a bounded, well-defined degradation prior space for the subsequent restoration process, where the quantized features are denoted as $f^{q}_{psf} = z^{q}$.

\noindent\textbf{Joint restoration learning.}
Concurrently, an image encoder $E_{img}$ extracts image features $f_{de}$ from the degraded input $I_{de}$.
A fusion module integrates $f_{de}$ with the quantized PSF features $f^q_{PSF}$, injecting the specific degradation information into the backbone. 
Finally, an image decoder $D_{img}$ reconstructs the clear image $\hat{I}_{c}$.

To optimize the latent space, we apply a codebook alignment loss $\mathcal{L}_{align}$ to realize the bidirectional optimization between the continuous features and the discrete priors~\cite{chen_mobile_2023}:
\begin{equation}
\label{eq:L_align}
\mathcal{L}_{align} = \Vert{sg[\hat{z}]-z^{q}}\Vert^2_2 + \beta\Vert{\hat{z}-sg[z^{q}]}\Vert^2_2,
\end{equation}
where $sg[\cdot]$ denotes the stop-gradient operation and $\beta$ is a commitment weighting hyperparameter. 
The first term updates the codebook $\mathcal{Z}$ to capture valid degradation patterns from the GT PSFs, while the second term constrains the encoder $E_{PSF}$ to prevent the continuous features from drifting away from the established discrete space. 

The entire network in Stage I is optimized end-to-end. The total training objective $\mathcal{L}_{Stage1}$ combines the image reconstruction loss $\mathcal{L}_{rec}$ (comprising the $L_1$ distance and a VGG-based perceptual loss) and the alignment loss $\mathcal{L}_{align}$:
\begin{equation}
\label{eq:L_stage1}
\mathcal{L}_{Stage1} = \mathcal{L}_{rec}(\hat{I}_{c}, I_{c}) + \mathcal{L}_{align}.
\end{equation}

Through this joint training on the large-scale AODLibpro dataset, the codebook $\mathcal{Z}$ is populated with diverse latent degradation patterns.
Meanwhile, the restoration network learns the optimal mapping from these discrete priors to clear images, establishing a robust foundation for the subsequent blind retrieval stage.

\subsection{Stage II: Blind Foundational Prediction}
\label{sec:stage2}
Stage I learns a discrete degradation space, but it requires GT PSFs that are unavailable in real-world blind scenarios. 
In Stage II, FoundCAC learns to retrieve discrete priors directly from degraded observations without GT PSFs.

\noindent\textbf{Prior retrieval via teacher-forcing.}
To preserve the learned degradation space, we strictly freeze the codebook $\mathcal{Z}$ and the pre-trained $E_{PSF}$ as the teacher model. 
We then introduce a blind PSF predictor $E_{pred}$. Given only the degraded image $I_{de}$, $E_{pred}$ predicts the continuous latent features $\hat{f}_{PSF}$. 
To constrain this ill-posed prediction, we employ a teacher-forcing strategy. 
The GT PSF map $I_{PSF}$ is encoded by the frozen $E_{PSF}$ and quantized by $\mathcal{Z}$ to produce the target discrete prior $f^q_{PSF}$. 
To force the predicted $\hat{f}_{PSF}$ to approximate $f^q_{PSF}$ before quantization, we formulate a feature matching loss $\mathcal{L}_{FM}$~\cite{chen2022real}. 

\noindent\textbf{Foundational blind correction.}
The predicted features $\hat{f}_{PSF}$ are quantized by the frozen codebook $\mathcal{Z}$ to obtain the discrete prior $\hat{f}^q_{PSF}$.
This prior is then integrated into the restoration network, initialized from Stage I, to reconstruct the clear image $\hat{I}_{c}$. 
The overall objective for Stage II is:
\begin{equation}
\label{eq:L_stage2}
\mathcal{L}_{Stage2} = \mathcal{L}_{rec}(\hat{I}_{c}, I_{c}) + \mathcal{L}_{FM}.
\end{equation}

By replacing unconstrained continuous regression with discrete prior retrieval, our paradigm effectively reduces the ambiguity of the highly ill-posed blind correction task. 

\subsection{Stage III: Target-Specific Adaptation}
\label{sec:stage3}

While Stage II provides robust zero-shot capabilities, specific target lenses often exhibit unique domain shifts. To further enhance restoration performance on unseen lenses, we jointly fine-tune $E_{pred}$ and the restoration network using target-domain image pairs and the reconstruction loss $\mathcal{L}_{rec}$. During this process, the codebook $\mathcal{Z}$ is strictly frozen to anchor the framework to the established physical priors.

This codebook-freezing strategy constrains $E_{pred}$ to map target aberrations into the LPR space, yielding dual benefits. 
In \textit{full-shot} scenarios, it serves as a powerful initialization to elevate the final restoration performance.
Under \textit{few-shot} conditions, it explicitly prevents overfitting, ensuring highly efficient adaptation to target distributions under limited data conditions.

\section{Experimental Results}
\label{sec:Experimental_Results}
\begin{table*}[!t]
    \begin{center}
    \vspace{-0.5em}
        \caption{Comparison with potential blind lens aberration correction pipelines on \textit{RealLens-Sim}. 
        We report the PSNR/SSIM/LPIPS results under each sub-test-lenses-set. 
        The latency of each method to process an image of $1920\times1280$ is also provided. 
        The \textbf{best} and \underline{second} results are highlighted.} 
        \vspace{-1em}
        \label{tab:main}

\resizebox{1.0\textwidth}{!}
{

\renewcommand{\arraystretch}{1.55}
\setlength{\tabcolsep}{1mm}{
\begin{tabular}{lcccccc}
\bottomrule[0.17em]
\multirow{2}{*}{\textbf{Method}} & \multicolumn{1}{l}{\multirow{2}{*}{\textbf{Latency (s)}}} & \multicolumn{5}{c}{\textit{RealLens-Sim}} \\ \cline{3-7} 
 & \multicolumn{1}{l}{} & \textbf{MOS-S/A} & \textbf{MOS-Meta} & \textbf{MA} & \textbf{High-end} & \textbf{Average} \\ \bottomrule[0.17em]
Fast two-step~\cite{eboli2022fast} &\textbf{0.390} &20.69/0.720/0.3127  &21.58/0.655/0.4488  &\underline{27.13}/0.767/0.1811  &{28.10}/0.827/0.1718  &24.38/0.742/0.2786  \\ \hline
Universal IR model (S3Diff)~\cite{zhang2024degradation} &9.933  &20.34/0.746/0.2615  &21.36/0.687/0.4229    &{26.90}/0.799/0.1779  &\textbf{29.69}/0.857/0.1423  &24.57/0.772/0.2512  \\ \hline
ZEBASELib-PT~\cite{gong2024physics} + SwinIR~\cite{liang2021swinir} &0.782  &23.09/0.791/0.2969  &18.14/0.679/0.4931  &25.91/0.837/0.1274  &28.16/0.901/0.1053  &23.83/0.802/0.2557  \\
ZernikeLib-PT~\cite{jiang2024computational} + SwinIR~\cite{liang2021swinir} &0.782  &24.41/0.822/0.1841  &20.50/0.707/0.3806  &25.16/0.850/0.1248  &27.23/0.898/\underline{0.0932}  &24.33/0.819/0.1957  \\
AODLib-LensNet-PT~\cite{cote2021deep} + SwinIR~\cite{liang2021swinir} &0.782  &23.10/0.796/0.2937  &18.56/0.678/0.4993  &25.97/0.858/0.1224  &27.49/\textbf{0.904}/0.0977  &23.78/0.809/0.2533  \\
AODLib-EAOD-PT~\cite{jiang2024flexible} + SwinIR~\cite{liang2021swinir} &0.782  &26.72/0.853/0.1597  &{21.96}/0.748/0.3671  &26.02/0.861/\underline{0.1015}  &27.72/0.903/0.0953  &25.14/0.839/0.1842  \\
\rowcolor{Gray}
\textbf{AODLibpro-PT (Ours)} + SwinIR~\cite{liang2021swinir} &0.782 &\underline{26.77}/\underline{0.856}/\underline{0.1610}  &\underline{22.46}/\underline{0.751}/\underline{0.3599}  &26.50/\textbf{0.861}/0.1014  &27.93/\textbf{0.903}/0.0948    &\underline{25.91}/\underline{0.843}/\underline{0.1793}  \\
\rowcolor{Gray}
\textbf{FoundCAC (Ours)} &\underline{0.417} &\textbf{27.23}/\textbf{0.866}/\textbf{0.1405}  &\textbf{23.95}/\textbf{0.771}/\textbf{0.3111}  &\textbf{27.54}/\textbf{0.866}/\textbf{0.0919}  &\underline{28.64}/\textbf{0.904}/\textbf{0.0904}  &\textbf{26.84}/\textbf{0.852}/\textbf{0.1585}  \\ \bottomrule[0.17em]
\end{tabular}
}
}
    \end{center}
    \vspace{-1.5em}
\end{table*}

\begin{figure*}[!t]
  \centering
  \includegraphics[width=1.0\textwidth]{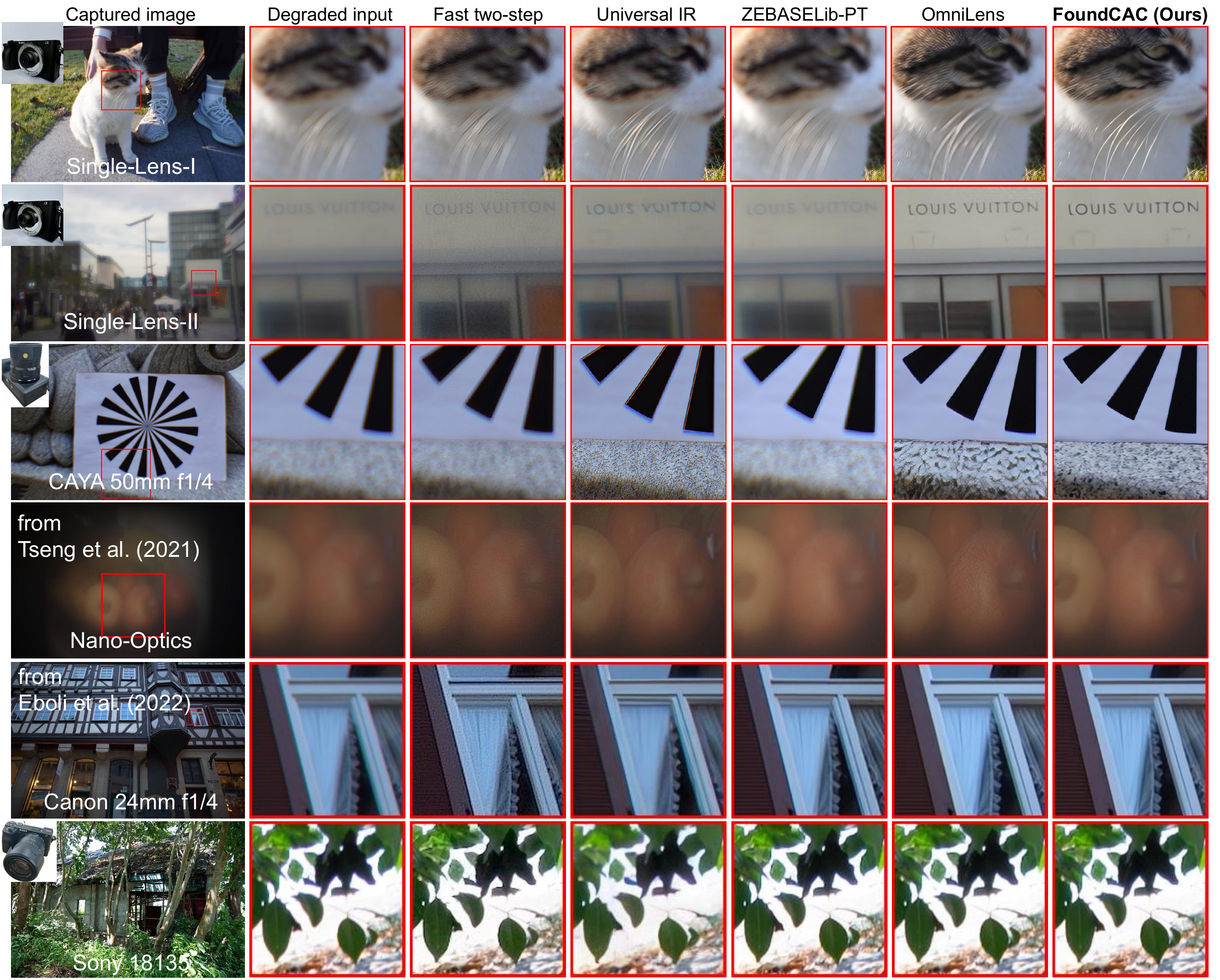}
  \vspace{-2em}
  \caption{Visual results of representative blind lens aberration correction methods on \textit{RealLens-Snap}.}
  \label{fig:visual}
  \vspace{-1.5em}
\end{figure*}

\subsection{Implementation Details}
\noindent\textbf{Datasets.}
We sample $m_1=200$ and $m_2=3$ instances per class for AODLibpro \texttt{train} and AODLibpro \texttt{test}, yielding $3,600$ training lenses and $54$ test lenses with no overlap. 
For AODLibpro \texttt{train}, we synthesize $40$ degraded images per lens from Flickr2K~\cite{timofte2017ntire} via a precise imaging simulator~\cite{yang2024end}. 
For AODLibpro \texttt{test}, the degraded images are synthesized based on additional $26$ collected clear images. 
Meanwhile, through manually designing and gathering open source designs, we construct a real-world lens aberration dataset \textit{RealLens-Sim}, consisting of $4$ minimalist optical systems with spherical or aspheric surfaces (MOS-A/S)~\cite{jiang2024flexible}, $1$ metalens from~\cite{tseng2021neural} (MOS-Meta), one smartphone lens under $3$ misalignment (MA) cases, and $2$ high-end lenses including an ultra-wide lens and a smartphone lens (High-end).
The same simulator is applied to simulate the paired images of them for numerical evaluation.
For real-world cases, we collect real-snapped optical degradation images \textit{RealLens-Snap} with fabricated single lenses, the commercial minimalist lens $CAYE$ $50mm$ $f1/4$, two DSLRs of $Canon$ $24mm$ $f1/4$ and $Sony$ $18135$, and nano-optics data from~\cite{tseng2021neural}. 

\noindent\textbf{Training details.}
The default FoundCAC employs a UNet-style architecture with $3$-group ResBlock encoder/decoders~\cite{chen2022real}, a discrete codebook of size 1024~\cite{chen2022real,esser2021taming}, and a deep bottleneck of $4$ Residual Swin Transformer Blocks (RSTB)~\cite{liang2021swinir}, providing a strong inductive bias for spatially-varying degradations~\cite{jiang2024minimalist}. 
Models are trained on the AODLibpro \texttt{train} set following our three-stage paradigm. 
Stage I jointly optimizes the VQ codebook and restoration network using GT PSFs for $200K$ iterations. 
Stage II freezes the codebook to train the blind PSF predictor via teacher-forcing for another $200K$ iterations. 
For Stage III, full-shot adaptation utilizes the complete DIV2K training set as target data~\cite{timofte2017ntire}, whereas few-shot adaptation fine-tunes the model on merely $5\%$ of the target data for $10K$ iterations with a reduced learning rate.
During training, we use $256\times256$ random crops with standard geometric augmentations and a batch size of $16$. We optimize via Adam~\cite{kingma2014adam}, with the learning rate cosine-annealed from $2\times 10^{-4}$ to $1\times 10^{-6}$. 
Training runs on two RTX 4090 GPUs, and inference latency is measured on a single RTX 4090. 
The subsequent sections evaluate blind restoration on representative real lenses (Sec.~\ref{sec:realens}), benchmark zero-shot generalizability (Sec.~\ref{sec:generalizability}), and investigate target-specific adaptation performance (Sec.~\ref{sec:adaptation}). Specific dataset configurations and evaluation protocols are detailed within each respective subsection.

\subsection{Blind Correction for RealLens Aberrations}
\label{sec:realens}
To evaluate zero-shot performance on existing optical designs, we assess blind restoration on the \textit{RealLens} dataset, using \textit{RealLens-Sim} for full-reference quantitative evaluation and \textit{RealLens-Snap} for real-capture qualitative evaluation.

\noindent\textbf{Numerical evaluation on \textit{RealLens-Sim}.} 
On \textit{RealLens-Sim}, we evaluate the overall capability of blind lens aberration correction methods to handle aberrations from real optical designs, as shown in Table~\ref{tab:main}.
The suite includes the state-of-the-art deconvolution method fast two-step~\cite{eboli2022fast}, a universal Image Restoration (IR) model for real-world degradations represented by S3Diff~\cite{zhang2024degradation} trained under the BSRGAN~\cite{zhang2021designing} data regime, and various LensLib-PT methods containing LensLibs of ZEBASELib~\cite{gong2024physics}, ZernikeLib~\cite{jiang2024computational}, AODLib-LensNet~\cite{cote2021deep}, and AODLib-EAOD in OmniLens~\cite{jiang2024flexible}. 
For the LensLib-PT methods, we adopt SwinIR~\cite{liang2021swinir} as the network architecture for its superior overall performance in OmniLens. 
In addition to our final FoundCAC trained on AODLibpro, we also report SwinIR trained on AODLibpro as a reference result.

Overall, our full framework achieves state-of-the-art blind aberration correction results under this real-design simulation protocol.
{\ourframework}~delivers pronounced gains on challenging aberration cases such as MOS-S/A and MOS-Meta, effectively handles highly stochastic misalignment aberrations, and surpasses fast two-step methods on high-end lens aberrations, which are specifically designed for them. 
Notably, a universal IR model trained without optical degradation data copes with the relatively mild aberrations of high-end lenses, yet struggles with more complex and severe cases, indicating the importance of dedicated research on lens aberration correction.
Finally, the last three rows show that, atop the OmniLens baseline, AODLibpro and LPR yield average PSNR improvements of $0.77$ dB and $0.93$ dB, respectively, verifying that both our data and model paradigm designs enhance the capacity of the LensLib-PT-based pipeline for blind aberration correction.
More visual results are provided in the appendix.

\noindent\textbf{Qualitative evaluation on \textit{RealLens-Snap}.}
Using real-snapped images (\textit{RealLens-Snap}), we further conduct qualitative validation of the representative methods in Table~\ref{tab:main}, as shown in Figure~\ref{fig:visual}. 
FoundCAC shows clear advantages in minimalist lenses, reflected in better handling of blur, stronger suppression of purple fringing, and little introduction of artifacts and ringing. 
On metalens images, although the model is trained only on refractive lenses, our method yields improved image clarity compared to baseline approaches.
Finally, similar to the fast two-step method~\cite{eboli2022fast} specialized for high-end optics, FoundCAC also improves the image quality of high-end DSLR lenses by enhancing sharpness and correcting purple fringing.
In summary, RealLens-Sim and RealLens-Snap provide complementary evidence for zero-shot generalization across real-design simulations and real-captured images.

\begin{table}[!t]
    \begin{center}
        \caption{Comparison with representative methods for blind aberration correction on the AODLibpro \texttt{test} benchmark. The \textbf{best} and \underline{second} results are highlighted. Red numbers in the gray rows denote the absolute gain over the corresponding baseline architectures.}
        \vspace{-1em}
        \label{tab:bench}
        \resizebox{1.0\linewidth}{!}{
        \renewcommand{\arraystretch}{1.1}
        \setlength{\tabcolsep}{0.5mm}{
            \begin{tabular}{lcccc}
            \toprule[0.17em]
            \textbf{Network} & \textbf{Latency (s)} & \textbf{PSNR$\uparrow$} & \textbf{SSIM$\uparrow$} & \textbf{LPIPS$\downarrow$} \\ 
            \midrule[0.17em]
            RRDBNet~\cite{wang2018esrgan} & 0.296 & 27.22 & 0.852 & 0.1519 \\
            MIMOUnet~\cite{cho2021rethinking} & 0.405 & 28.11 & \underline{0.874} & 0.1651 \\
            Restormer~\cite{zamir2022restormer} & 1.859 & 27.11 & 0.867 & 0.1430 \\
            X-Restormer~\cite{chen2024comparative} & 2.797 & 28.11 & 0.870 & 0.1408 \\
            PromptIR~\cite{potlapalli2023promptir} & 2.063 & 27.14 & 0.869 & 0.1404 \\
            DiffBIR~\cite{lin2024diffbir} & 66.130 & 27.52 & 0.833 & 0.1430 \\
            S3Diff~\cite{zhang2024degradation} & 9.933 & 23.10 & 0.762 & 0.1678 \\
            FOVKPN~\cite{chen2021extreme_quality} & \underline{0.166} & 27.23 & 0.851 & 0.1586 \\
            DFUnet~\cite{chen2021optical} & \textbf{0.137} & 27.04 & 0.841 & 0.1639 \\ \hline
            NAFNet~\cite{chen2022simple} & 0.353 & 27.86 & 0.868 & 0.1382 \\
            SwinIR~\cite{liang2021swinir} & 0.782 & 28.29 & 0.871 & 0.1346 \\
            UNet-RSTB (Baseline) & 0.383  & 28.46 & 0.873 & 0.1318 \\ \hline
            \rowcolor{Gray}
            \textbf{NAFNet~\cite{chen2022simple} + LPR} & 0.374 & 28.07~{\scriptsize \color{red}(+0.21)} & 0.873 & \underline{0.1308} \\
            \rowcolor{Gray}
            \textbf{SwinIR~\cite{liang2021swinir} + LPR} & 0.846  & \underline{28.70}~{\scriptsize \color{red}(+0.41)} & \underline{0.874} & 0.1318 \\ 
            \rowcolor{Gray}
            \textbf{UNet-RSTB + LPR (FoundCAC)} & 0.417 & \textbf{29.12}~{\scriptsize \color{red}(+0.66)} & \textbf{0.877} & \textbf{0.1191} \\ 
            \bottomrule[0.17em]
            \end{tabular}
        }
        }
    \end{center}
    \vspace{-2em}
\end{table}

\begin{table*}[!t]
    \begin{center}
        \caption{Quantitative evaluation of adaptation strategies on simulated test sets (PSNR [dB] $\uparrow$ / LPIPS $\downarrow$). ``Scratch'' denotes training from random initialization. FoundCAC pretraining consistently gives the best results and shows clear advantages under few-shot adaptation.}
        \vspace{-1em}
        \label{tab:adaptation_large}
        \resizebox{1.0\linewidth}{!}{
        \renewcommand{\arraystretch}{1.2}
        \setlength{\tabcolsep}{1.8mm}{
            \begin{tabular}{lcccccc}
            \toprule[0.17em]
            \multirow{3}{*}{\textbf{Training Strategy}} & \multicolumn{4}{c}{\textbf{Regime I: Depth-Independent}} & \multicolumn{2}{c}{\textbf{Regime II: Depth-Aware}} \\
            \cmidrule(lr){2-5} \cmidrule(l){6-7}
            & \multicolumn{2}{c}{Single-Lens-I} & \multicolumn{2}{c}{Smartphone Lens} & \multicolumn{2}{c}{Single-Lens-I} \\
            \cmidrule(lr){2-3} \cmidrule(lr){4-5} \cmidrule(l){6-7}
            & Full-shot & Few-shot & Full-shot & Few-shot & Full-shot & Few-shot \\ 
            \midrule[0.17em]
            FoundCAC (Zero-shot) & \multicolumn{2}{c}{27.52 / 0.126} & \multicolumn{2}{c}{28.72 / 0.089} & \multicolumn{2}{c}{25.78 / 0.213} \\ \hline
            Scratch \textit{w./o.} PSF & 31.66 / 0.117 & 27.22 / 0.195 & 33.04 / 0.063 & 32.07 / 0.091 & 26.63 / 0.203 & 22.16 / 0.322 \\
            Scratch \textit{w./} GT PSF & 31.86 / 0.120 & 28.09 / 0.190 & 33.03 / 0.066 & 32.02 / 0.089 & 26.36 / 0.200 & 23.69 / 0.285 \\ \hline
            \rowcolor{Gray}
            \textbf{FoundCAC pretraining (Ours)} & \textbf{31.94} / \textbf{0.110} & \textbf{30.90} / \textbf{0.118} & \textbf{34.08} / \textbf{0.059} & \textbf{33.34} / \textbf{0.065} & \textbf{26.75} / \textbf{0.183} & \textbf{26.25} / \textbf{0.200} \\
            \bottomrule[0.17em]
            \end{tabular}
        }
        }
    \end{center}
    \vspace{-2.0em}
\end{table*}

\begin{figure*}[!t]
  \centering
  \includegraphics[width=0.99\textwidth]{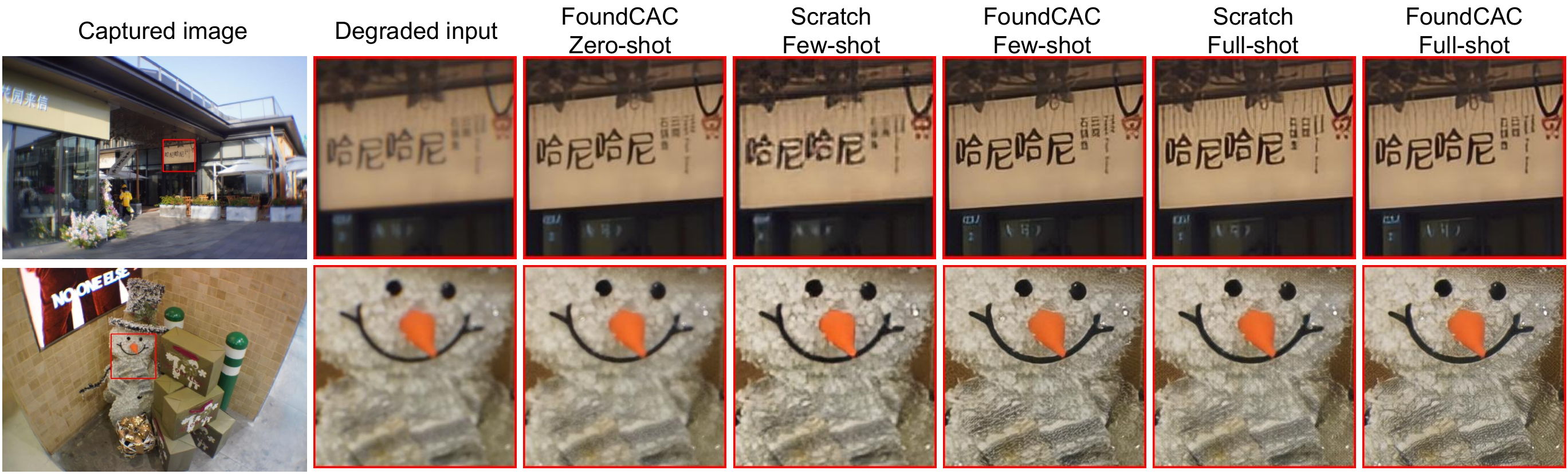}
  \vspace{-0.5em}
    \caption{Visual comparison on real captured images for Single-Lens-I. Under few-shot adaptation, our method recovers clearer structures and textures than scratch training. Under full-shot adaptation, it also produces slightly cleaner results.}
  \label{fig:target adaptation}
  \vspace{-1.5em}
\end{figure*}

\subsection{Performance and Generalization}
\label{sec:generalizability}
To evaluate the zero-shot generalization of FoundCAC and the applicability of the LPR paradigm across different network architectures, we benchmark all methods on unseen data. All baseline models and our framework are trained on AODLibpro \texttt{train} and evaluated on AODLibpro \texttt{test}.

\noindent\textbf{Comparison with state-of-the-art aberration correction networks.} 
The compared methods cover diverse architectures: the classic super-resolution networks RRDBNet~\cite{wang2018esrgan} and SwinIR~\cite{liang2021swinir}; the general image restoration models MIMOUNet~\cite{cho2021rethinking}, NAFNet~\cite{chen2022simple}, Restormer~\cite{zamir2022restormer}, X-Restormer~\cite{chen2024comparative}, and PromptIR~\cite{potlapalli2023promptir}; the diffusion-based models DiffBIR~\cite{lin2024diffbir} and S3Diff~\cite{zhang2024degradation}; and the CAC-specific designs FOVKPN~\cite{chen2021extreme_quality} and DFUnet~\cite{chen2021optical}. 
As reported in Table~\ref{tab:bench}, the default FoundCAC achieves state-of-the-art results on the benchmark. 
PromptIR shows limited improvement over its baseline Restormer, suggesting that standard prompt mechanisms are insufficient to capture complex, spatially-varying optical degradations. 
Furthermore, diffusion-based methods also show lower performance and higher inference latency.
These results suggest that, for lens aberration correction, explicitly learned optical priors are more effective than generic natural image priors.
By incorporating degradation priors derived from physical optics, FoundCAC also outperforms domain-specific designs such as FOVKPN and DFUnet.
\noindent\textbf{Generalizability of the LPR-guided training paradigm.}
A key property of our method is that the proposed multi-stage training paradigm is not restricted to a specific backbone. 
When applied to different architectures, the same training procedure can be used to learn the LPR and the corresponding blind restoration model.
To verify this, we apply the LPR-guided training strategy to representative baselines, including SwinIR~\cite{liang2021swinir} and NAFNet~\cite{chen2022simple}.
Specifically, we train these models using the same discrete prior learning and blind prediction stages.
As shown in Table~\ref{tab:bench}, models trained under the LPR-guided paradigm consistently outperform their standard end-to-end counterparts, without noticeable inference overhead.
This result indicates that explicitly modeling discrete optical priors improves blind restoration across different backbone architectures.

\subsection{Adaptation Efficiency: From Full-Shot to Few-Shot}
\label{sec:adaptation}

We evaluate target-specific adaptation under two degradation regimes: depth-independent and depth-aware. For the depth-independent regime, we evaluate single-lens-I and smartphone lenses on \textit{RealLens-Sim}. For the depth-aware regime, we evaluate depth-coupled single-lens-I on the DIV2K validation set with simulated depth-aware degradation~\cite{yang2025efficient} and on real-world \textit{RealLens-Snap} images. For each regime, we compare two adaptation protocols: full-shot and few-shot. In the few-shot protocol, only $5\%$ of the target data and fine-tuning iterations used in full-shot adaptation are retained. Since full-shot adaptation remains costly for each target lens, we examine whether the pretrained prior learned from our optical library improves adaptation efficiency under limited target data.
We employ the same codebook-freezing strategy for both full-shot and few-shot adaptations.
Table~\ref{tab:adaptation_large} shows that, across both degradation regimes and both adaptation protocols, initializing from FoundCAC consistently outperforms training from scratch. This indicates that the pretrained LPR provides a transferable prior for target-specific adaptation across different lenses and degradation types.

\noindent\textbf{Full-shot adaptation.}
With sufficient target data, FoundCAC still provides a better initialization than scratch training and leads to slightly better final performance. In the depth-independent regime, our method improves PSNR from $33.04$ dB to $34.08$ dB on the smartphone lens. A similar trend is observed in the depth-aware regime, where our method reaches $26.75$ dB, compared with $26.63$ dB for scratch training. These results indicate that the pretrained LPR remains useful even when the target data are sufficient.

\noindent\textbf{Few-shot adaptation.}
The advantage of pretraining becomes more evident under a limited adaptation budget. Under few-shot adaptation, scratch training degrades substantially because the model must learn target-specific degradation representations from very limited data. 
As a result, our method improves PSNR from $27.22$ dB to $30.90$ dB on the standard single-lens setting. 
In the depth-aware regime, scratch training achieves $22.16$ dB, and continuous GT PSF supervision reaches $23.69$ dB, whereas our method achieves $26.25$ dB. This result is close to the full-shot performance of $26.75$ dB while using only $5\%$ of the target data and optimization steps.

\noindent\textbf{Visual results on real-world captures.}
We further apply the models adapted in the depth-aware regime to real-world \textit{RealLens-Snap} images for qualitative evaluation. 
As shown in Figure~\ref{fig:target adaptation}, the visual results are consistent with the quantitative results on simulated data. Under few-shot adaptation, scratch-trained models still exhibit residual blur and artifacts, while our method recovers clearer structures and textures. 
Under full-shot adaptation, the visual difference is smaller, but our method still produces slightly cleaner restorations in some regions. 
These observations suggest that the adaptation benefit of FoundCAC is not limited to simulation and can transfer to real captured images.

\begin{figure*}[!t]
  \centering
  \includegraphics[width=0.99\textwidth]{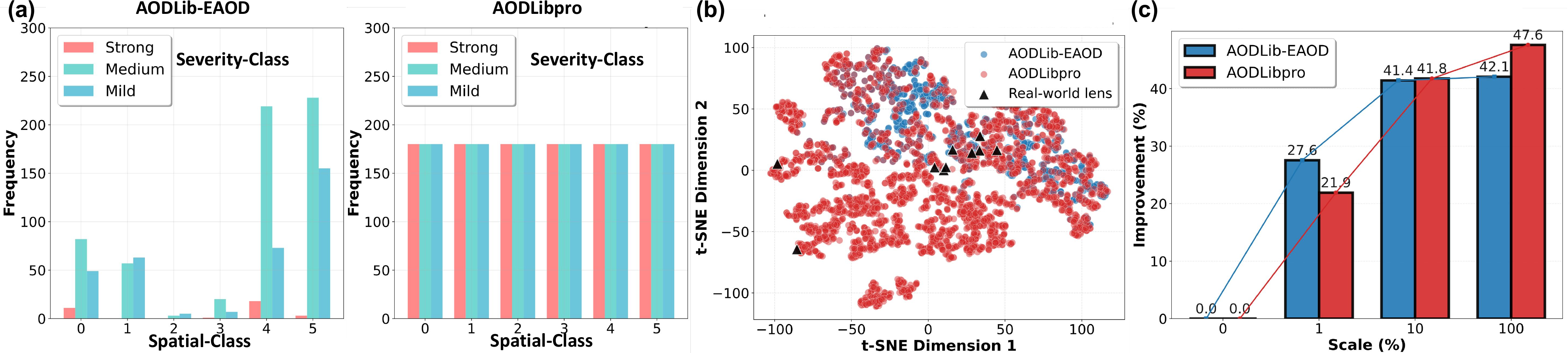}
  \caption{AODLibpro \textit{v.s.} AODLib-EAOD in terms of uniformity of aberration distributions, coverage over real-world lenses, and scalability. 
  (a) Histogram of degradation type distributions in the sampled lenses. 
  (b) LensLib coverage visualization based on OIQ evaluated per FoV and wavelength. 
  (c) Improvements of SwinIR trained with LensLibs of different scales over the method without LensLib~\cite{eboli2022fast}. 
  The improvement is averaged across the PSNR and LPIPS.}
  \label{fig:ablib}
  \vspace{-1em}
\end{figure*}

\begin{table}[!t]
    \begin{center}
        \caption{Ablations on AODLibpro construction strategies.} 
        \vspace{-1em}
        \label{tab:ab_aodlibpro}
        \resizebox{0.99\linewidth}{!}{ 
        \renewcommand{\arraystretch}{1.2}
        \setlength{\tabcolsep}{4mm}{
            \begin{tabular}{lcc}
            \toprule[0.17em]
            \textbf{Design Specification} & \textbf{PSNR $\uparrow$} & \textbf{LPIPS $\downarrow$} \\ 
            \midrule[0.17em]
            Baseline (OmniLens configuration) & 25.14 & 0.1842 \\
            + Aspheric Surface & 25.72 & 0.1893 \\
            \rowcolor{Gray}
            \textbf{+ Aspheric Surface \& Image Plane Perturbation (Ours)} & \textbf{25.91} & \textbf{0.1793} \\ 
            \midrule[0.17em]
            \textbf{Sampling Basis} & \textbf{PSNR $\uparrow$} & \textbf{LPIPS $\downarrow$} \\ 
            \midrule[0.17em]
            RMS & 25.14 & 0.1842 \\
            Severity-Class & \textbf{26.10} & 0.1859 \\
            Spatial-Class & 25.85 & 0.1815 \\
            \rowcolor{Gray}
            \textbf{Hybrid (Ours)} & 25.91 & \textbf{0.1793} \\ 
            \bottomrule[0.17em]
            \end{tabular}
        }
        }
    \end{center}
    \vspace{-0.5em}
\end{table}

\subsection{Ablation Studies}
\label{sec:ablations}
We perform ablations to investigate the individual effectiveness of AODLibpro and the multi-stage training paradigm.
To ensure fairness, any experiment that modifies the training data is tested on \textit{RealLens-Sim} to assess the overall framework, while models trained on the full AODLibpro \texttt{train} are evaluated on the AODLibpro \texttt{test} to evaluate the model paradigm.

\noindent\textbf{Evaluation of AODLibpro.} 
We train a SwinIR model~\cite{liang2021swinir} under different LensLib settings to evaluate their impact.
Table~\ref{tab:ab_aodlibpro} shows that incorporating additional design specifications improves blind correction performance.
Furthermore, regarding the sampling strategy, compared with the RMS-based sampling basis, the proposed Severity-Class and Spatial‑Class lead to better generalization, and combining them contributes to a more comprehensive improvement. 
We further demonstrate the advantages of AODLibpro over its baseline AODLib‑EAOD in Figure~\ref{fig:ablib}.
Benefiting from specification expansion and the hybrid sampling basis, the samples in AODLibpro are uniformly distributed across degradation severity and spatial variation patterns (Figure~\ref{fig:ablib}~(a)), while the overall aberration distribution is broader and can cover all lens samples in \textit{RealLens-Sim} (Figure~\ref{fig:ablib}~(b)). 
These properties also improve the scalability of AODLibpro, where increasing the data scale brings consistent gains (Figure~\ref{fig:ablib}~(c)), in contrast to AODLib-EAOD.

\begin{table}[!t]
    \begin{center}
        \caption{Ablation study on prior representations and training paradigms.} 
        \vspace{-1em}
        \label{tab:ablation_mechanisms}
        \resizebox{1.0\linewidth}{!}{
        \renewcommand{\arraystretch}{1.2}
        \setlength{\tabcolsep}{1.5mm}{ 
            \begin{tabular}{lllccc}
            \toprule[0.17em]
            & \textbf{Model Variant} & \textbf{Prior Repr.} & \textbf{PSNR $\uparrow$} & \textbf{SSIM $\uparrow$} & \textbf{LPIPS $\downarrow$} \\ 
            \midrule[0.17em]
            (a) & Baseline (Standard Blind) & None & 28.46 & 0.873 & 0.1318 \\
            (b) & Baseline (Extended to 400k) & None & 28.46 & 0.875 & 0.1237 \\
            (c) & Direct PSF Feature Prediction & Continuous & 28.42 & 0.867 & 0.1287 \\
            (d) & Direct GT-PSF Fusion & Continuous & 28.73 & 0.873 & 0.1269 \\ \hline
            (e) & FoundCAC Stage I (GT-Guided) & Discrete VQ & 28.97 & 0.874 & 0.1194 \\ 
            \rowcolor{Gray}
            (f) & \textbf{FoundCAC Stage II (Full Blind)} & \textbf{Discrete VQ} & \textbf{29.12} & \textbf{0.877} & \textbf{0.1191} \\ 
            \bottomrule[0.17em]
            \end{tabular}
        }
        }
    \end{center}
    \vspace{-1.0em}
\end{table}

\noindent\textbf{Analysis on the role of training stages.}
We evaluate the proposed LPR-guided multi-stage training paradigm in Table~\ref{tab:ablation_mechanisms}.
To ensure fair comparison, we extend the baseline to 400k iterations (Model~b) to match our total training budget. Its PSNR remains saturated at $28.46$ dB, confirming a fundamental capacity bottleneck in purely data-driven models.
Next, predicting continuous PSF features (Model~c) reduces PSNR to $28.42$ dB, suggesting that unconstrained continuous spaces introduce ambiguity and noisy prior estimates. 
Even fusing GT continuous features (Model~d) yields limited improvement ($28.73$ dB), indicating insufficient structural constraints. 
In contrast, the GT-guided discrete VQ representation (Model~e) improves PSNR to $28.97$ dB. This shows that the codebook serves as a degradation prior by converting continuous regression into discrete retrieval. 
Finally, the full multi-stage paradigm (Model~f) achieves $29.12$ dB. By freezing the dictionary learned in Stage I, Stage II restricts blind correction to discrete prior retrieval while jointly optimizing the predictor and restoration backbone.
This synergy explains our empirical observation that the final blind model is capable of outperforming the Stage I GT-guided result under certain backbone architectures, firmly supporting the benefit of stage-wise separation.

\section{Conclusion}
\label{sec:conclusion}
We present FoundCAC as a foundational framework for blind aberration correction. 
Built on the broad coverage and scalability of AODLibpro, FoundCAC adopts a multi-stage training paradigm to learn a latent PSF representation.
This design enables the use of discrete physical priors while maintaining fully blind inference. 
Extensive experiments demonstrate that FoundCAC achieves superior zero-shot performance under complementary evaluation on synthetic LensLib, real-design simulations, and real-captured images, while providing an efficient adaptation strategy for specific unseen lenses.
It achieves performance comparable to specialized methods on mild aberrations in high-end DSLR lenses, while showing clear advantages on more severe degradations in minimalist optical systems.

\noindent\textbf{Limitations and future research directions.} 
For real-captured images, reliable paired full-reference evaluation remains difficult because reference-lens swapping may change geometry, ISP/color response, sensor noise, and depth-dependent blur.
Moreover, the LPR codebook is learned from synthetic refractive LensLib data, so its coverage may be limited for fabricated lenses with manufacturing errors, assembly tolerances, sensor/ISP interactions, or diffractive effects.
Given the scalability of AODLibpro, an immediate next step is to expand the physical prior library to include diffractive optical elements and more complex metasurfaces, with the goal of improving zero-shot generalization further.
Another direction is to extend FoundCAC to compound physical degradations, where optical aberrations are coupled with depth of field~\cite{abuolaim2020defocus,ruan2022learning,yang2023aberration,yang2025efficient}, under-display interference~\cite{feng2021removing,wang2024perspective}, low-light conditions~\cite{liu2023low}, and sensor noise~\cite{zheng2025darkdiff}. 
Finally, although the proposed target-specific adaptation is data-efficient, unsupervised on-device adaptation remains an important direction for improving practical deployment.

\ifpeerreview \else
\section*{Acknowledgments}
This research was funded by the Natural Science Foundation of Zhejiang Province (Grant No. LZ24F050003), the National Natural Science Foundation of China (Grant No. 62473139), the Hunan Provincial Research and Development Project (Grant No. 2025QK3019), and the opening project of the State Key Laboratory of Autonomous Intelligent Unmanned Systems (Grant No. ZZKF2025-2-10).
\fi

\bibliographystyle{IEEEtran}
\bibliography{references}

\begin{figure*}[!h]
  \centering
  \includegraphics[width=1.0\textwidth]{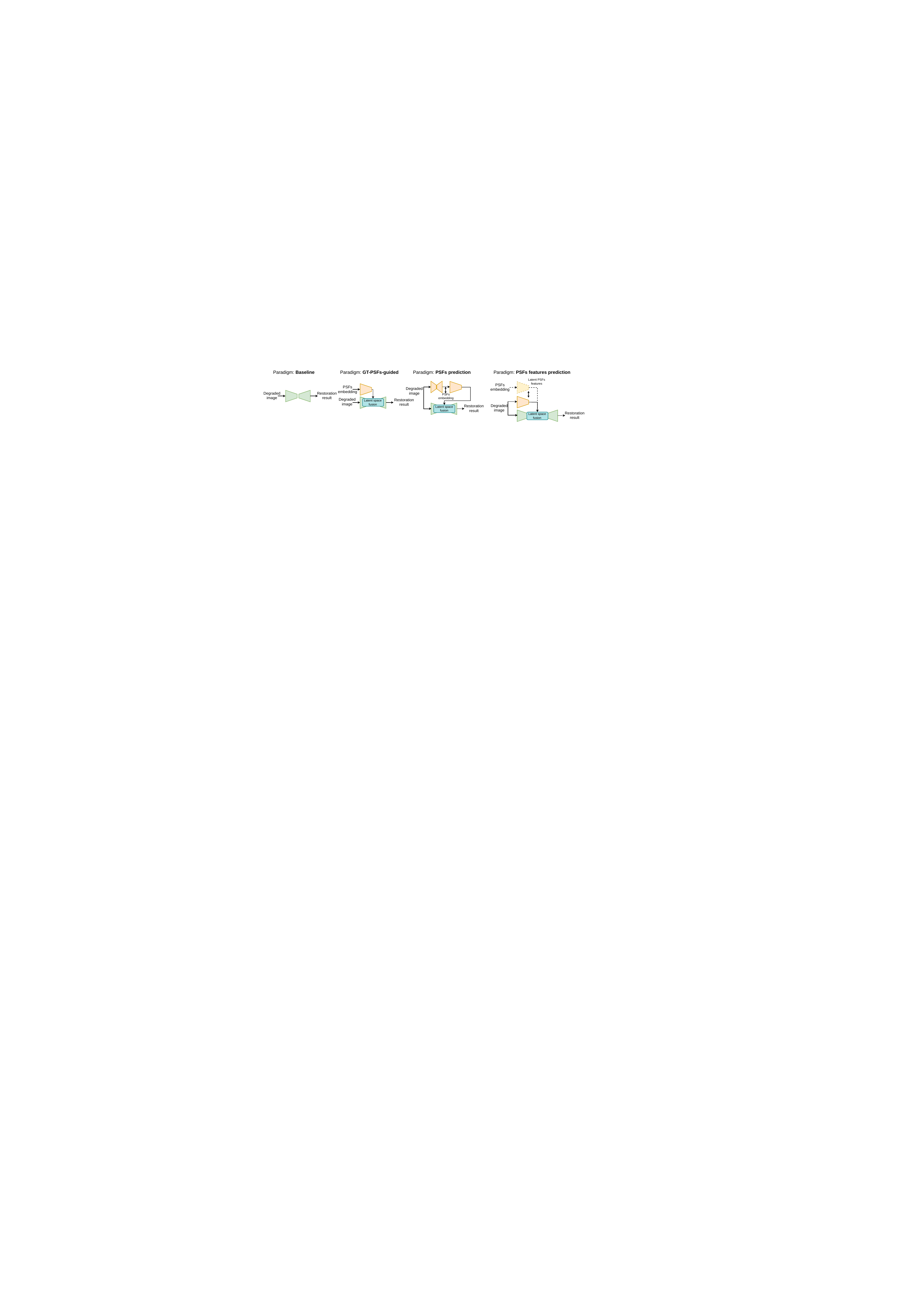}
  \caption{Illustration of the model paradigms in motivation.}
  \label{fig:motimodel}
\end{figure*}

\section{Detailed Settings for Experiments in Motivation}
\label{sup:moti}
Figure~\ref{fig:motimodel} illustrates the schematic diagrams of the different model paradigms compared in Table~1 of the main manuscript. 
To ensure a fair comparison, all baseline architectures share the identical backbone design as the default FoundCAC. 
Furthermore, the feature fusion mechanisms between the PSF condition and image features (\textit{i.e.}, channel-wise concatenation), as well as the bottleneck architecture, are kept strictly consistent with the default FoundCAC framework. For paradigms involving the prediction of continuous PSFs or latent PSF features, an $L_1$ loss is applied for direct supervision. All other training hyperparameters and settings strictly follow those of the blind prediction stage in FoundCAC. 

Finally, to objectively benchmark network performance, all models discussed in this section are trained from scratch on the AODLibpro-Train set and evaluated on the AODLibpro-Test benchmark under identical environments.

\section{Details for AODLibpro Construction}
\label{sup:oiq}
\subsection{Details for supplemented specifications}
\noindent\textbf{Definition of the aspheric surface.}
The aspheric lens surface is an optical surface whose curvature deviates from a constant-radius sphere. 
Unlike conventional spherical surfaces, its profile is mathematically defined by higher-order polynomials, enabling precise control over light refraction across the entire aperture.
The height of a standard aspheric surface~\cite{sun2021end} is defined as a function of the radial distance $r$:
\begin{equation}
\label{eq:appendix1}
    h(r) = \frac{cr^2}{1+\sqrt{1-(1+\kappa)c^2r^2}}+\sum^{N_A}_{i=2}a_{2i}r^{2i},
\end{equation}
where $c$ denotes the curvature, $\kappa$ is the conic coefficient, $a_{2i}$'s are higher-order coefficients, and $N_A$ defines the highest-order aspheric coefficient.

\noindent\textbf{Image distance perturbation constrained by depth of field.}
We constrain the image distance perturbation amplitude within acceptable limits using the Depth of Field (DoF) formula:
\begin{equation}
\label{eq:appendix2}
   \Delta L = \Delta L_1+\Delta L_2=\frac{F\delta L^2}{f^2+F\delta L} +\frac{F\delta L^2}{f^2-F\delta L},
\end{equation}
where $\delta$ is the permissible circle of confusion diameter, $f$ is the lens focal length, $F$ is the F-number, $L$ is the image distance, $\Delta L_1$ is the near DOF, and $\Delta L_2$ is the far DOF. Image distance perturbation range is constrained within $[-\Delta L_1, \Delta L_2]$, and $\delta$ is set to $24 \mu m$ in this work.
The perturbation probability $\gamma$ is set to $25\%$ empirically.

\subsection{Details for hybrid sampling basis}
\noindent\textbf{Calculation of OIQ.}
Following~\cite{qian2026unicac}, OIQ incorporates traditional fidelity-based image quality metrics (PSNR and SSIM~\cite{wang2004image}) as well as the SFR-based metric to provide an image quality assessment considering optical properties:
\begin{equation}
    \small
    OIQ = \lambda_{1}~\frac{PSNR}{50}  + \lambda_{2}~\frac{SSIM - 0.5}{0.5} + \lambda_{3}~OIQE,
    \label{eq:Score}
\end{equation}
where PSNR and SSIM are processed following~\cite{liang2024ntire} to obtain normalized metrics, while SFR is represented as OIQE~\cite{jiang2024minimalist}, which denotes the ratio of the SFR of the evaluated target to that of a lens without optical degradation. 
Consistent with~\cite{qian2026unicac}, the weights are set to $\lambda_1=0.4,\lambda_2=0.3,\lambda_3=0.3$ to balance the contributions of the three normalized metrics. 

\begin{figure*}[!h]
  \centering
  \includegraphics[width=0.7\textwidth]{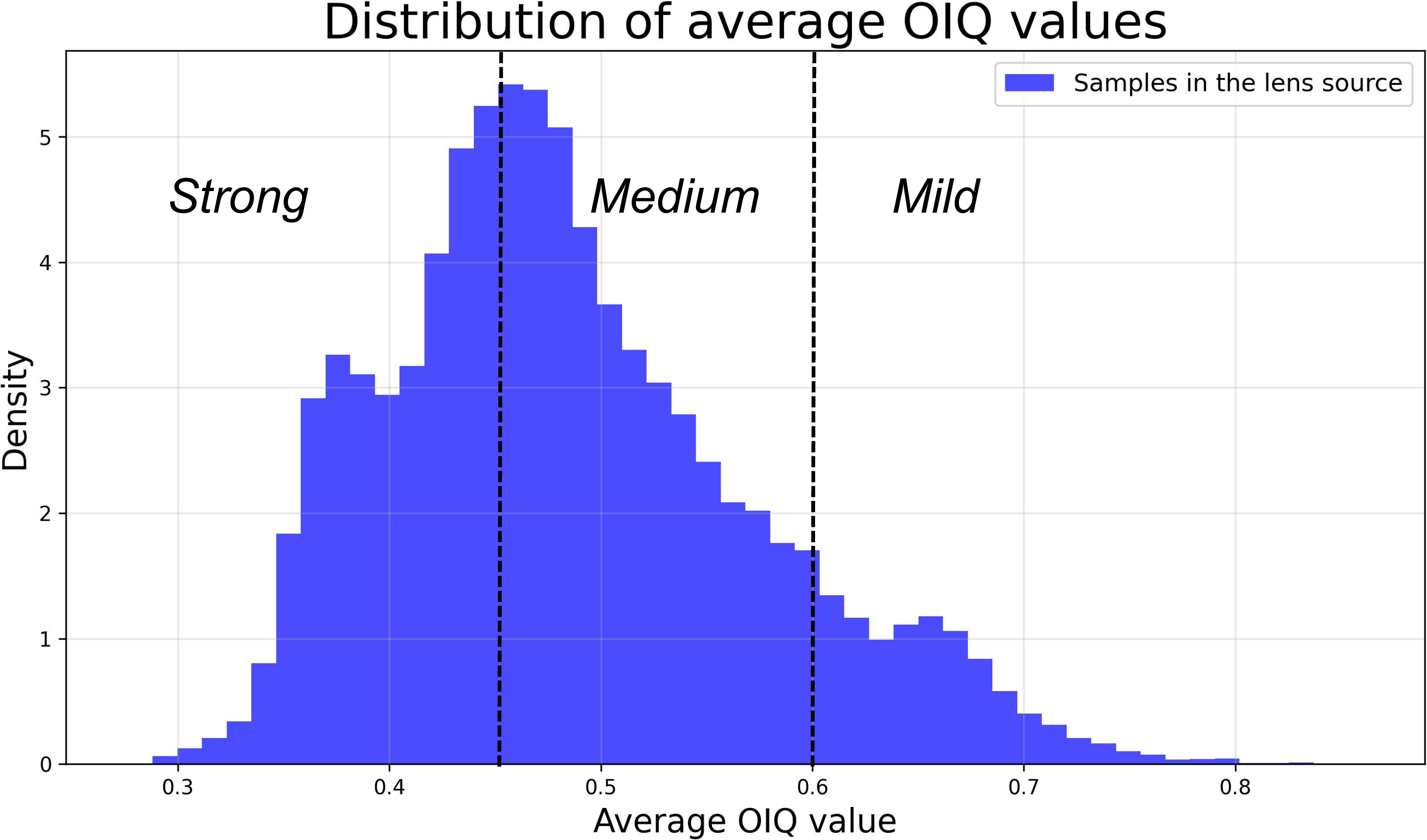}
  \caption{Illustration of the definition for Severity-Class.}
  \label{fig:oiqrange}
\end{figure*}

\noindent\textbf{Definition of Severity-Class.}
The average OIQ across the $5$ knife-edge image patches of different FoVs is calculated to represent the overall severity level of the target lens's optical degradation.
The average OIQ lies within $[0, 1]$, where a larger value indicates lower overall optical degradation severity (higher optical image quality). 
Therefore, we divide the range into $3$ intervals to categorize different optical degradation severity levels for sampling, as shown in Figure~\ref{fig:oiqrange}.

\noindent\textbf{Calculation of the spatial uniformity.}
As proposed in~\cite{qian2026unicac}, we use the coefficient of variation (CV) from the variance and mean of OIQ across the $5$ FoVs, then use it to compute a spatial uniformity metric $U_{S}$ that measures the uniformity of the optical degradation spatial distribution:
\begin{equation}
    CV= \frac{Std(\left \{ OIQ_{i} | i=1,...,5 \right \}  )}{Avg(\left \{ OIQ_{i} | i=1,...,5 \right \})},
    \label{eq:CV}
\end{equation}
\begin{equation}
    U_{{S}}=e^{-5 CV},
    \label{eq:uniformity fov}
\end{equation}
where $OIQ_{i}$ is the OIQ of the $i_{th}$ FoV.
A larger $U_S$ indicates that the optical degradation spatial distribution of the lens is more uniform.

\begin{figure*}[!h]
  \centering
  \includegraphics[width=1.0\textwidth]{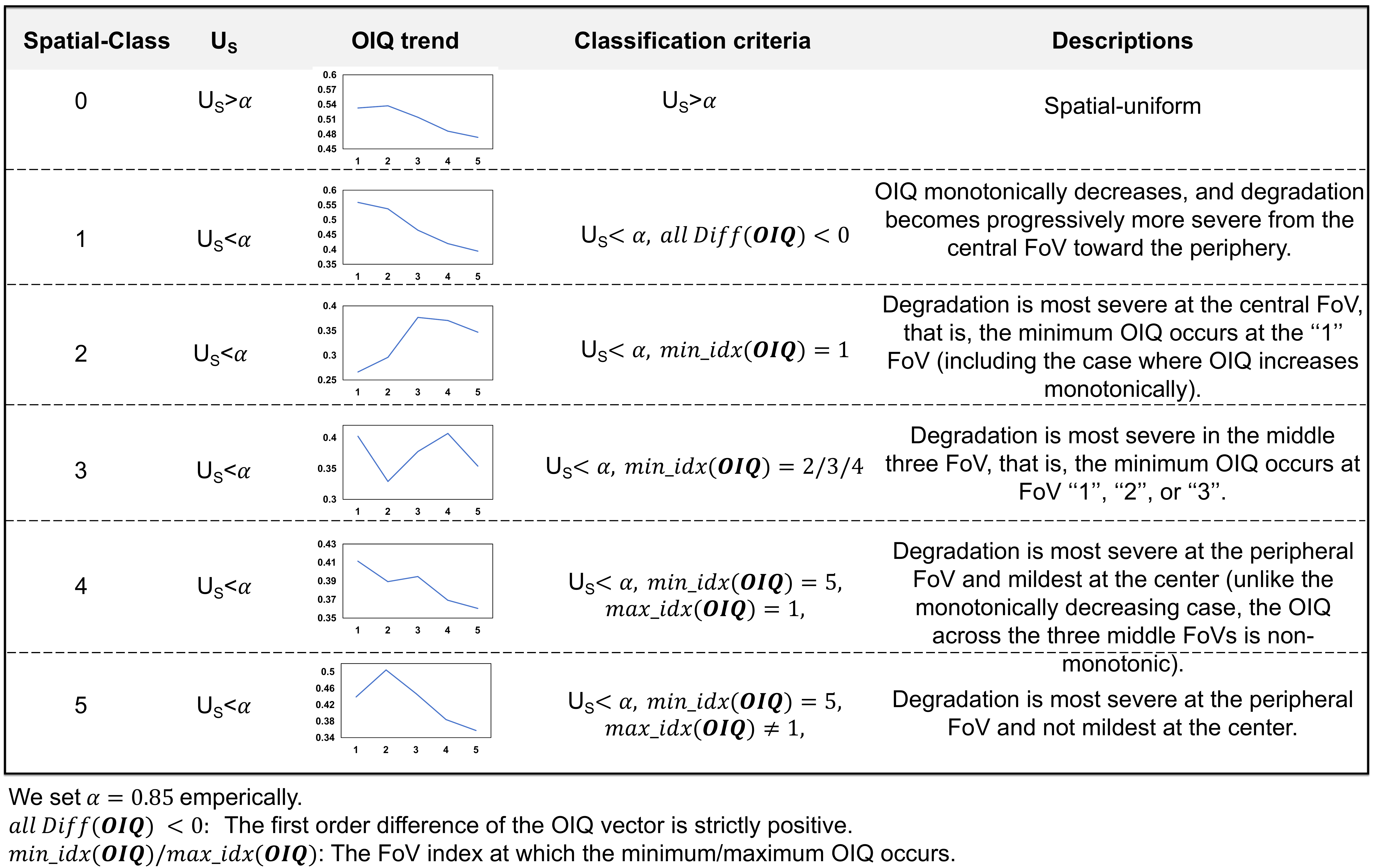}
  \caption{Illustration of the definition of Spatial-Class. Notably, for cases where degradation is most severe at the peripheral FoV, we define Spatial-Classes ``4'' and ``5'' because we observe that some lenses, despite showing the most severe degradation at the peripheral FoV, also exhibit relatively severe degradation at the center FoV. This degradation pattern clearly differs from the typical degradation pattern of a sharp center and blurred edge, which should be independently considered.}
  \label{fig:odclass}
\end{figure*}

\noindent\textbf{Definition of Spatial-Class.}
Figure~\ref{fig:odclass} provides detailed definitions for each Spatial‑Class, including the $U_S$ value range, schematic OIQ trend plots, classification criteria, and supplementary descriptions. This basis allows any OIQ to fall into one class, enabling the classification and description of all possible optical degradation patterns. To the best of our knowledge, there is currently no such detailed categorization of optical degradation distribution patterns, so we make a preliminary attempt to explore this problem here. 
Table 5 in the manuscript shows that using the proposed Spatial‑Class for sampling can construct a more effective LensLib.
We hope this classification approach can provide new insights for this field to understand optical degradation distribution patterns.

\begin{figure*}[!h]
  \centering
  \includegraphics[width=0.8\textwidth]{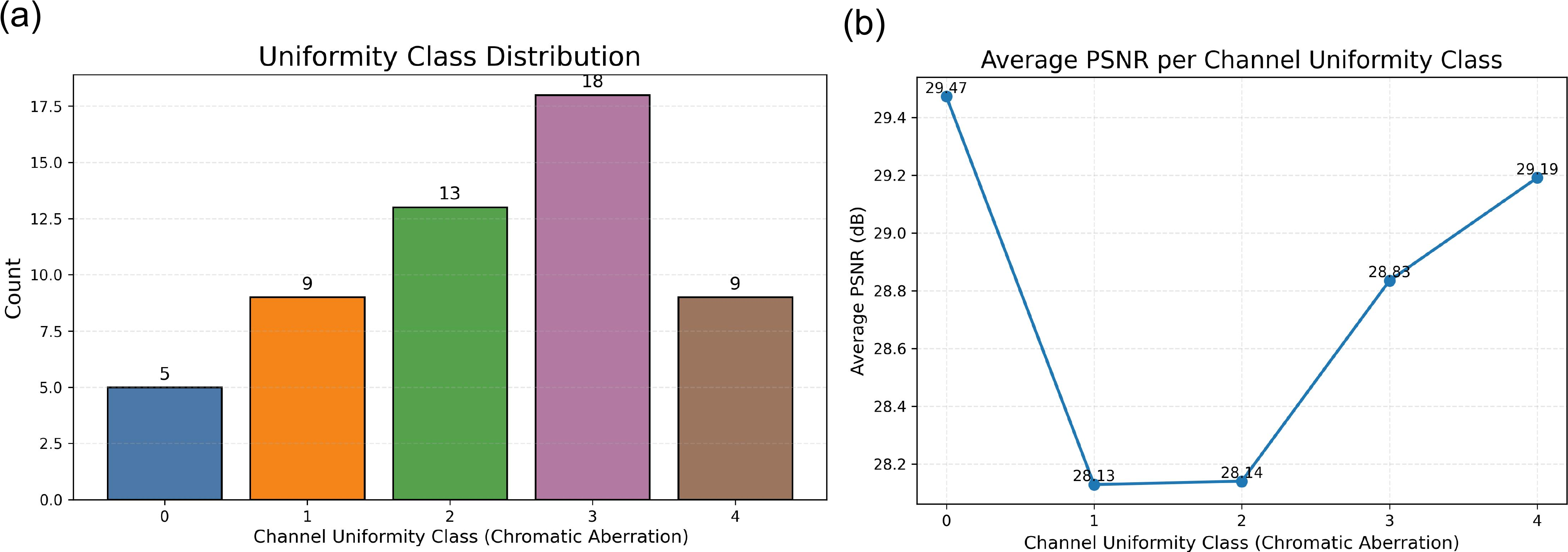}
  \caption{Evidence for omitting chromatic aberration as the optical degradation classification criteria. (a) Distribution of different levels of chromatic aberrations in AODLibpro \texttt{Test}. (b) Performance (PSNR) of FoundCAC under different chromatic aberration levels.}
  \label{fig:chromatic}
\end{figure*}

\noindent\textbf{Discussion on chromatic aberrations.}
We do not include chromatic aberration as a criterion in optical degradation classification because preliminary experiments show that it has little impact on the final correction results, as shown in Figure~\ref{fig:chromatic}. 
Using the same computation pipeline as $U_S$, we compute the channel-wise uniformity of OIQ to quantify the severity of chromatic aberration and divide it into $5$ categories, where a smaller category index indicates more severe chromatic aberration. 
The distribution of chromatic aberration in AODLibpro \texttt{Test} is shown in Figure~\ref{fig:chromatic} (a). 
We tally the performance of our trained FoundCAC at different chromatic levels and find no obvious pattern in its performance as the severity of chromatic aberration changes, as shown in Figure~\ref{fig:chromatic} (b). 
This indicates that chromatic aberration has little effect on aberration correction models trained on LensLib, even though it is an often-discussed optical degradation pattern. 
Meanwhile, our experiments also include cases with pronounced chromatic aberration, as in Figure 4 in the main manuscript, where our method effectively suppresses purple fringing caused by chromatic aberration.
These evidences indicate that although chromatic aberration is not considered in data sampling, it does not affect the model’s performance in this respect.

\begin{figure*}[!h]
  \centering
  \includegraphics[width=0.6\textwidth]{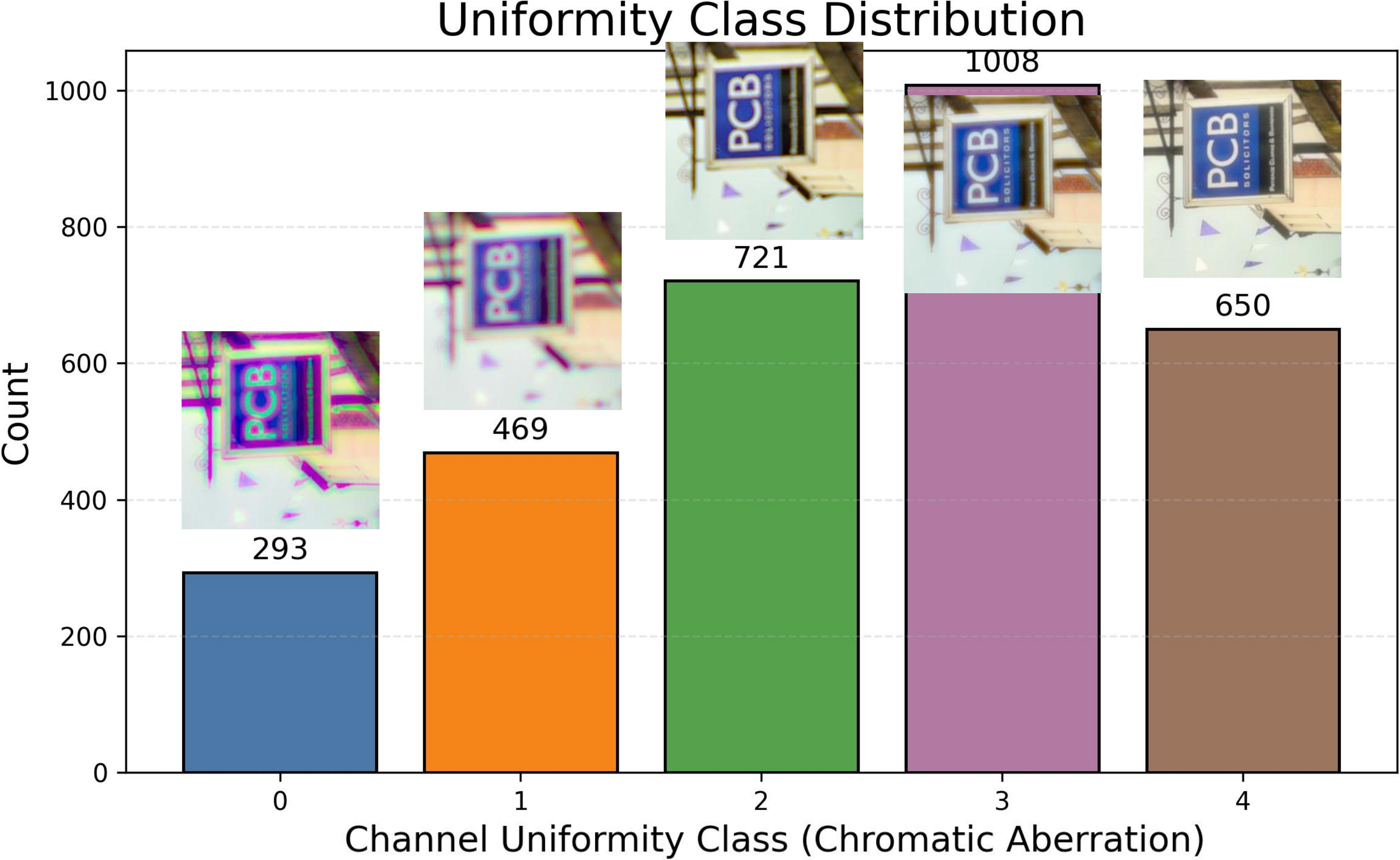}
  \caption{{Distribution of different chromatic aberration levels in AODLibpro \texttt{Train}. Visualization cases for each level are also provided.}}
  \label{fig:chromatic_train}
\end{figure*}

{Then, because the configuration of the EAOD algorithm does not include settings for a cemented doublet structure, the optimization imposes no strong constraint on chromatic aberration, which causes many generated samples to exhibit noticeable chromatic aberration. 
Figure~\ref{fig:chromatic_train} shows the distribution of chromatic aberration severity in our AODLibpro \texttt{Train}, from which it can be seen that a considerable portion of the training samples reveal obvious chromatic aberration. We believe that these chromatically degraded samples in the training set endow the model with the ability to handle chromatic aberration, enabling it to effectively address purple fringing across the test cases.}

\section{Detailed Settings for Main Experiments}
\label{sup:setting}
\subsection{Details on imaging simulation}
\label{sup:simulation}
For AODLibpro \texttt{Train}, AODLibpro \texttt{Test}, and \textit{RealLens-Sim}, we obtain the paired optical degradation images corresponding to each lens design via imaging simulation.
The imaging simulator from DeepLens~\cite{yang2024end} is adopted, considering its precise calculation of PSFs and simulation of optical degradation images by patch-wise convolution. 
Specifically, we feed the design files of each sample in these LensLibs (in Zemax or parameter table format) into DeepLens for ray tracing to compute PSFs at $64$ fields of view and $31$ sampled wavelengths in the visible band. 
We match the closest sensor from the sensor library ($4\mu{m}-2K$, $8\mu{m}-2K$, $12\mu{m}-2K$, $16\mu{m}-2K$) based on the image height, discretize and sample the PSFs according to the pixel size, and stack them across RGB channels according to the wavelength response characteristics, ultimately obtaining the PSF arrays for each FoV and channel used in the simulation.
It is worth noting that we applied equivalent downsampling to sensor resolution and pixel size to facilitate the use of large‑scale public $2K$ high‑quality image datasets. 
Meanwhile, unifying the resolution also helps to systematically control variables to build a benchmark for exploring the aberration correction task.
Finally, the computed PSFs are used to convert clear images into the corresponding optical degradation images via patch-wise convolution, while the pipeline also accounts for sensor ISP and noise, as in most imaging simulation workflows~\cite{chen2021extreme_quality,chen2021optical}. For each lens in the $3$ datasets, we will open source its Zemax design file, the computed PSF array, and the simulated paired optical degradation images.

\subsection{Motivation for setting AODLibpro \texttt{Test}}
Previous studies on blind aberration correction have applied specific test data~\cite{eboli2022fast,gong2024physics,jiang2024flexible}, which makes it inconvenient to evaluate the performance of model paradigms. 
In this situation, the optical degradation distributions of test lenses vary widely and unpredictably across works, so evaluations often measure the joint effect of training data and the model paradigm rather than the paradigm’s own ability to learn optical degradation from data. 
Therefore, with training fixed on AODLibpro \texttt{Train}, we propose constructing AODLibpro \texttt{Test} as a benchmark for evaluating aberration correction networks. 
This benchmark has the following advantages: i) lenses in \texttt{Train} and \texttt{Test} are non‑overlapping samples drawn from the same EAOD‑generated lens source, which ensures that optical degradation distributions in \texttt{Test} are independently unseen while keeping the domain gap moderate, thereby focusing on the paradigm’s ability to learn optical degradation; ii) constructing \texttt{Test} also uses the hybrid sampling basis, yielding optical degradation distributions that are uniform across spatial variation patterns and severity, with no data bias, thus enabling reliable evaluation of comprehensive aberration correction performance. 
We believe this benchmark setup can promote exploration of model paradigms and help ensure that the validations of various model designs are effective. 

\begin{figure*}[!h]
  \centering
  \includegraphics[width=0.99\textwidth]{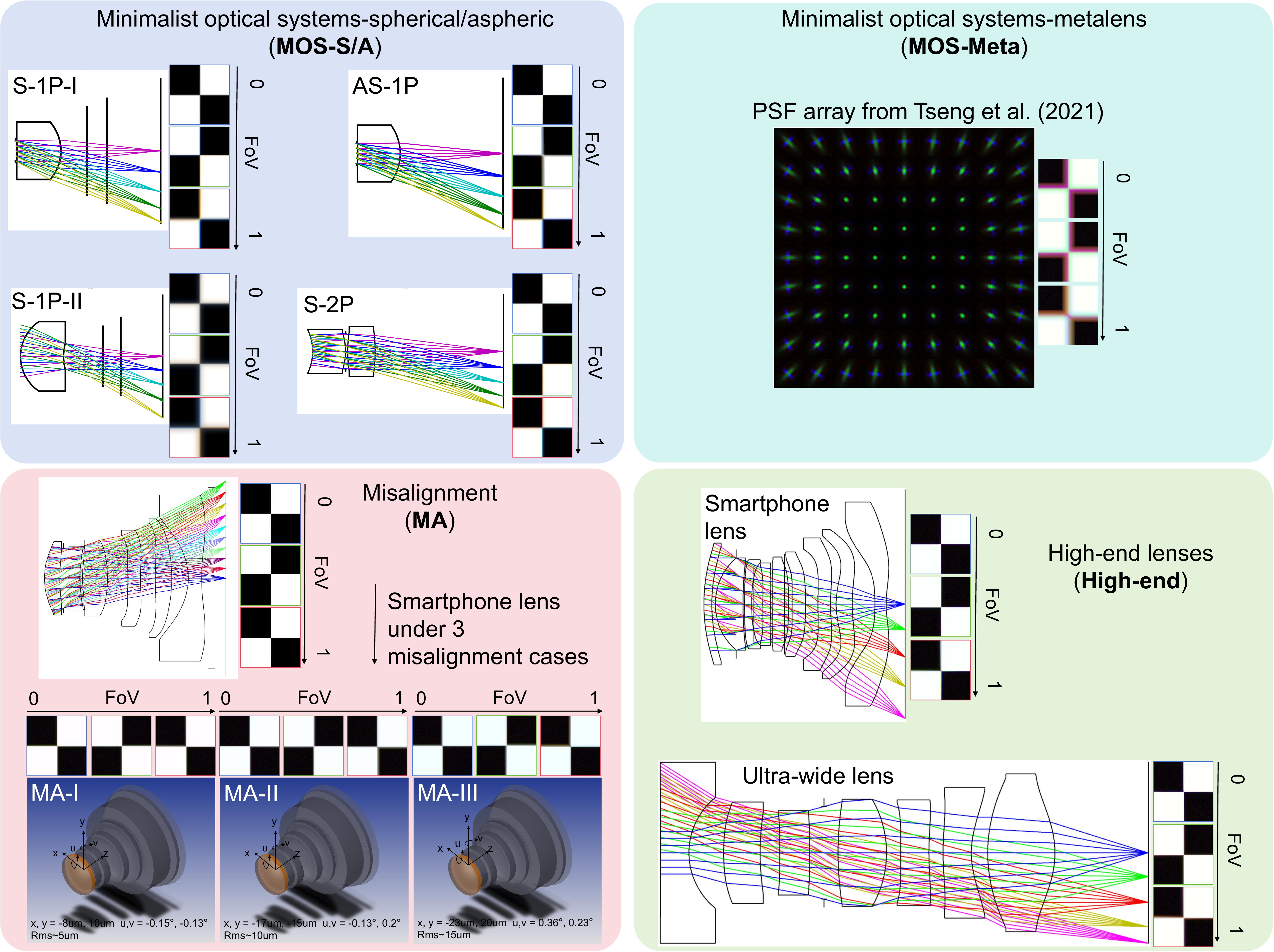}
  \caption{Illustration of lens designs, settings, and optical degradation patterns for \textit{RealLens-Sim}.}
  \label{fig:realsim}
\end{figure*}

\subsection{Illustration of lens designs in \textit{RealLens-Sim}}
Figure~\ref{fig:realsim} shows the structures of the test lenses used in \textit{RealLens‑Sim} and example imaging results. 
The lenses are sourced from open-source designs in the literature or from designs manually created by optical designers based on specifications in public patents or the needs of minimalist applications. 
In addition, we consider lens misalignment, a common but rarely addressed real-world factor that induces optical degradation. 
This typically occurs when the PSF size under the nominal lens design is below one pixel, that is, no optical degradation, but the decentering and tilt errors in manufacturing and assembly lead to random unknown optical degradation in the final imaging results. 
We select a smartphone lens whose original design yields no sensor-sampled optical degradation, set $3$ groups of random decentering and tilt errors with increasing magnitude within its tolerance range, and then perform ray tracing to compute PSFs for imaging simulation.

{Unlike AODLibpro \texttt{Test}, which provides a comprehensive evaluation in terms of optical degradation severity and spatial variation patterns in the imaging results, \textit{RealLens‑Sim} aims to provide test data from the perspective of lens types across real-world application scenarios, reflecting the practicality of the overall blind aberration correction framework and assessing the combined performance of training data and model paradigm. 
Admittedly, lens design cases in the real world are innumerable, and many are not open source, so one cannot include them all in the tests. Nevertheless, the lens types, application scenarios, and optical degradation distribution types covered by \textit{RealLens‑Sim} are the broadest among known open-source data, allowing it to serve as a strong evaluation benchmark for assessing the generalization of blind aberration correction methods.
In addition, because these lenses are manually designed by optical designers, they are out of domain relative to the LensLib generated by AOD methods, which can provide a fairer evaluation setting.}

\begin{figure*}[!h]
  \centering
  \includegraphics[width=0.99\textwidth]{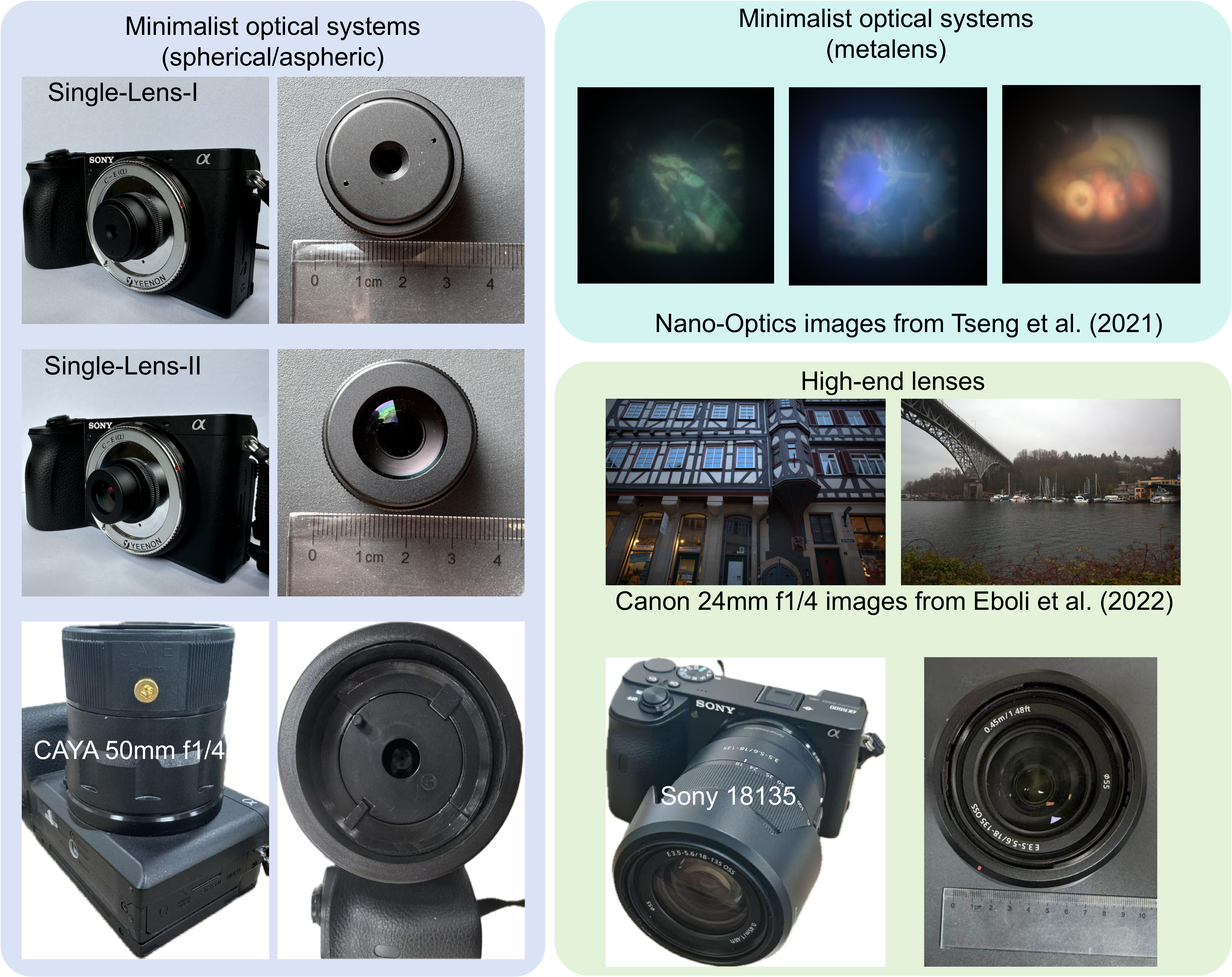}
  \caption{Schematic of the real snap setups for each lens in \textit{RealLens-Snap}. For all devices whose setups are shown, we use images snapped in real-world scenes, while for the others we directly use their open-source images~\cite{tseng2021neural,eboli2022fast}.}
  \label{fig:snap}
\end{figure*}

\subsection{Illustration of capture details for \textit{RealLens-Snap}}
We show the imaging setups used for our real-world captures in \textit{RealLens-Snap} in Figure~\ref{fig:snap}. 
In addition, the nano-optics data come from $3$ open-source images in~\cite{tseng2021neural}, and the $Canon$ $4mm$ $f1/4$ data come from two open-source images in~\cite{eboli2022fast}. 
Quantitative evaluation on \textit{RealLens-Snap} directly reflects the potential of blind aberration correction to improve image quality on real-world terminals. 
Given the difficulty of collecting real-world lenses under various applications for shooting, we strive to construct \textit{RealLens-Snap} covering commonly used scenarios for blind aberration correction, such as minimalist optical systems, high-end photographic equipment, and metalens imaging. 
The selected systems exhibit distinct optical degradation distributions, enabling a more comprehensive evaluation of the representative methods. 
To the best of our knowledge, we are the first work of blind aberration correction whose evaluation simultaneously covers minimalist optical systems with severe aberrations and high‑end lenses with mild aberrations.
We also hope to continuously collect more real snapped images with optical degradation in future work to broaden application scenarios.

\begin{table*}[!t]
    \begin{center}
        \caption{{Per‑lens results for competing blind lens aberration correction methods on \textit{RealLens-Sim}.}}
        \label{tab:main_sup}
        \resizebox{1.0\textwidth}{!}
{

\renewcommand{\arraystretch}{1.55}
\setlength{\tabcolsep}{1mm}{
\begin{tabular}{lccccccccccccccc}
\bottomrule[0.17em]
\multirow{2}{*}{\textbf{Method}} & \multicolumn{3}{c}{\textbf{S-1P-I}} & \multicolumn{3}{c}{\textbf{S-1P-II}} & \multicolumn{3}{c}{\textbf{AS-1P}} & \multicolumn{3}{c}{\textbf{S-2P}} & \multicolumn{3}{c}{\textbf{Nano-Optics}} \\
 & \textbf{PSNR} & \textbf{SSIM} & \textbf{LPIPS} & \textbf{PSNR} & \textbf{SSIM} & \textbf{LPIPS} &\textbf{ PSNR} & \textbf{SSIM} & \textbf{LPIPS} & \textbf{PSNR} & \textbf{SSIM }& \textbf{LPIPS} & \textbf{PSNR} & \textbf{SSIM} &\textbf{ LPIPS} \\ \hline
Fast two-step & 20.48 & 0.747 & 0.2783 & 18.87 & 0.681 & 0.4133 & 22.50 & 0.696 & 0.3077 & 20.90 & 0.758 & 0.2514 & 21.58 & 0.655 & 0.4488 \\
Universal IR & 20.05 & 0.761 & 0.2670 & 18.92 & 0.726 & 0.2698 & 22.00 & 0.734 & 0.2742 & 20.39 & 0.765 & 0.2351 & 21.36 & 0.687 & 0.4229 \\
ZEBASELib-PT & 23.84 & 0.813 & 0.2695 & 21.59 & 0.752 & 0.3736 & 22.26 & 0.767 & 0.3077 & 24.67 & 0.831 & 0.2369 & 18.14 & 0.679 & 0.4931 \\
ZernikeLib-PT & 26.39 & 0.855 & 0.1548 & 23.01 & 0.791 & 0.2534 & 23.61 & 0.779 & 0.1840 & 24.62 & 0.865 & 0.1442 & 20.50 & 0.707 & 0.3806 \\
AODLib-LensNet-PT & 23.70 & 0.826 & 0.2572 & 21.30 & 0.751 & 0.3978 & 22.48 & 0.768 & 0.3002 & 24.91 & 0.841 & 0.2194 & 18.56 & 0.678 & 0.4993 \\
AODLib-EAOD-PT & 27.83 & 0.880 & 0.1354 & 25.77 & 0.832 & 0.2002 & 24.29 & 0.803 & 0.1828 & 29.00 & 0.898 & 0.1202 & 20.10 & 0.740 & 0.3805 \\

\rowcolor{Gray}
AODLibpro-PT + SwinIR & 27.32 & 0.877 & 0.1450 & 25.06 & 0.828 & 0.2107 & 26.12 & 0.823 & 0.1662 & 28.59 & 0.894 & 0.1220 & 22.46 & 0.751 & 0.3599 \\
\rowcolor{Gray}
\textbf{FoundCAC (Ours)} & 27.52 & 0.880 & 0.1264 & 25.72 & 0.842 & 0.1827 & 27.90 & 0.895 & 0.1179 & 27.77 & 0.848 & 0.1349 & 23.95 & 0.771 & 0.3111 \\ \hline\hline
\multirow{2}{*}{\textbf{Method}} & \multicolumn{3}{c}{\textbf{MA-I}} & \multicolumn{3}{c}{\textbf{MA-II}} & \multicolumn{3}{c}{\textbf{MA-III}} & \multicolumn{3}{c}{\textbf{Ultra-wide lens}} & \multicolumn{3}{c}{\textbf{Smartphone lens}} \\ 
 & \textbf{PSNR} & \textbf{SSIM} & \textbf{LPIPS} & \textbf{PSNR} & \textbf{SSIM} & \textbf{LPIPS} &\textbf{ PSNR} & \textbf{SSIM} & \textbf{LPIPS} & \textbf{PSNR} & \textbf{SSIM }& \textbf{LPIPS} & \textbf{PSNR} & \textbf{SSIM} &\textbf{ LPIPS} \\ \hline
Fast two-step & 28.05 & 0.789 & 0.1708 & 26.54 & 0.742 & 0.1796 & 26.81 & 0.771 & 0.1930 & 28.23 & 0.828 & 0.1710 & 27.97 & 0.825 & 0.1725 \\
Universal IR & 27.96 & 0.822 & 0.1699 & 27.40 & 0.805 & 0.1776 & 25.34 & 0.770 & 0.1864 & 29.33 & 0.851 & 0.1464 & 30.06 & 0.864 & 0.1383 \\
ZEBASELib-PT & 26.66 & 0.855 & 0.1071 & 26.53 & 0.846 & 0.1232 & 24.55 & 0.809 & 0.1518 & 27.80 & 0.898 & 0.1123 & 28.52 & 0.905 & 0.0982 \\ 
ZernikeLib-PT & 26.47 & 0.881 & 0.1130 & 25.02 & 0.833 & 0.1173 & 24.01 & 0.835 & 0.1440 & 26.58 & 0.897 & 0.0929 & 27.87 & 0.900 & 0.0934 \\
AODLib-LensNet-PT & 27.12 & 0.886 & 0.1024 & 25.90 & 0.840 & 0.1185 & 24.87 & 0.847 & 0.1464 & 27.33 & 0.901 & 0.1041 & 27.66 & 0.908 & 0.0912 \\

AODLib-EAOD-PT & 26.80 & 0.892 & 0.0939 & 26.41 & 0.855 & 0.0932 & 24.86 & 0.836 & 0.1174 & 27.64 & 0.901 & 0.0968 & 27.80 & 0.906 & 0.0938 \\
\rowcolor{Gray}
AODLibpro-PT + SwinIR & 27.50 & 0.891 & 0.0934 & 26.45 & 0.851 & 0.0961 & 25.53 & 0.841 & 0.1148 & 27.93 & 0.900 & 0.0959 & 27.92 & 0.906 & 0.0936 \\
\rowcolor{Gray}
\textbf{FoundCAC (Ours)} & 28.80 & 0.897 & 0.0850 & 27.53 & 0.865 & 0.0861 & 26.28 & 0.838 & 0.1046 & 28.55 & 0.901 & 0.0915 & 28.72 & 0.905 & 0.0892\\\bottomrule[0.17em]
\end{tabular}
}
}
    \end{center}
\end{table*}

\begin{figure*}[!t]
  \centering
  \includegraphics[width=0.85\textwidth]{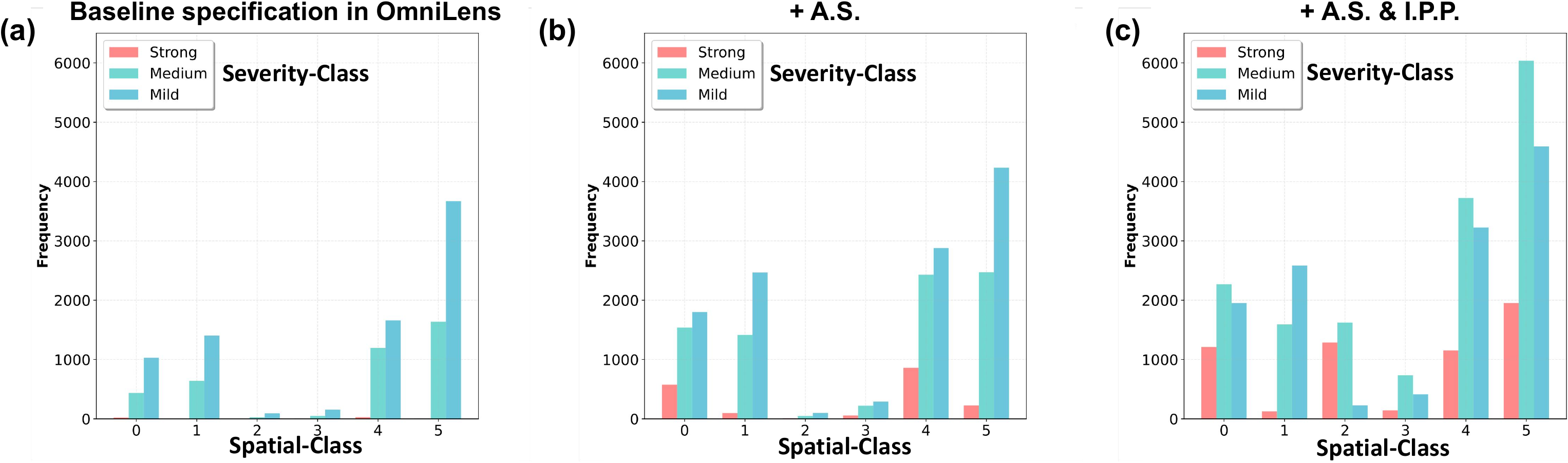}
  \caption{Optical degradation distributions of the lens source generated under different specification settings.}
  \label{fig:spe_od}
\end{figure*}

\subsection{Details for competing blind aberration correction methods}
For the fast two-step method~\cite{eboli2022fast}, since we are dealing with spatially varying optical degradation, we process the optical degradation images using $256\times256$ tiles with $128$ overlap, and keep all other settings the same as the defaults in its open‑source code. 
Regarding the choice of an open‑source, pretrained universal IR model for comparison, we consider it a primary option for users without an optical background to handle unknown optical degradation, because such methods aim to use highly generalizable large models to address real‑world unknown degradations. 
We directly load its pretrained weights~\cite{zhang2024degradation} to process our optical degradation data. 
For all LensLib‑PT methods~\cite{gong2024physics,jiang2024computational,cote2021deep,jiang2024flexible} on the data side, we generate paired optical degradation images using the same pipeline as AODLibpro based on each LensLib’s PSF arrays. 
Considering the inconsistency in the number of lenses across different Lenslibs, to ensure a fair comparison, we keep the total number of training images identical by changing the number of GT images degraded per lens. 
For the method in~\cite{gong2024physics}, although the proposed model paradigm is insightful, we use only its LensLib ZEBASELib to train SwinIR for comparison rather than the proposed model because it is not open-source. 
For the pipelines in~\cite{jiang2024computational} and~\cite{cote2021deep}, since they provide only the ideas for constructing a LensLib and do not involve aberration correction model design, we likewise use only the LensLibs built following their ideas and use SwinIR as the model.
For the SwinIR used as the network, we adopt the same $\mathcal{L}_{rec}$ as FoundCAC (L1 loss and perceptual loss), with a batch size of $16$ and $200K$ training iterations.

\subsection{Details for training competing aberration correction networks}
For all networks compared in Table~3 of the main manuscript, we use their architectures and retrain them on AODLibpro \texttt{Train}.
To fully exploit the capability of each method, we adopt the official training configurations.
Moreover, to ensure fairness in perceptual metrics, all methods are additionally trained with perceptual loss.

\begin{figure*}[!t]
  \centering
  \includegraphics[width=0.8\textwidth]{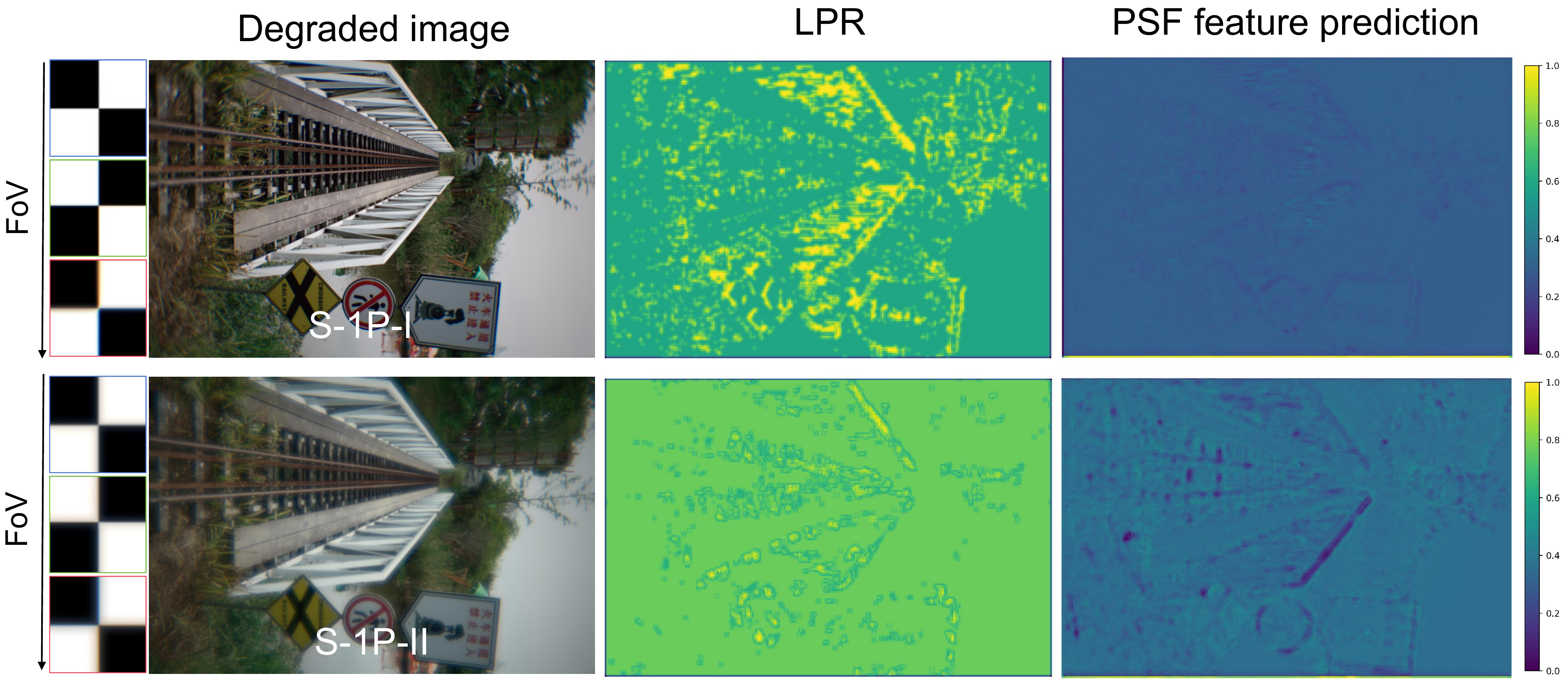}
  \caption{Visualization of the predicted latent PSF features. We use the PSF feature map before fusion to visualize the guidance of different PSF representations. The input optical degradation image and an example of its optical degradation distribution are shown in the first row.}
  \label{fig:vis_lpr}
\end{figure*}

\begin{table*}[!t]
    \begin{center}
        \caption{{{Per‑lens numerical evaluation for competing blind lens aberration correction methods on \textit{RealLens-Snap}.}} {The \textbf{first}, \underline{second}, and \textit{third} results are highlighted.}}
        \label{tab:nriqa}

\resizebox{1.0\textwidth}{!}
{

\renewcommand{\arraystretch}{1.55}
\setlength{\tabcolsep}{1mm}{
\begin{tabular}{c|cccc|cccc|cccc}
\bottomrule[0.17em]
\multirow{2}{*}{\textbf{Method}} & \multicolumn{4}{c|}{\textbf{Single-Lens-I}} & \multicolumn{4}{c|}{\textbf{Single-Lens-II}} & \multicolumn{4}{c}{\textbf{CAYA 50mm f1/4}} \\
 & \textbf{CLIPIQA$\uparrow$} & \textbf{NIQE$\downarrow$} & \textbf{MANIQA$\uparrow$} & \textbf{QAlign$\uparrow$} & \textbf{CLIPIQA$\uparrow$} & \textbf{NIQE$\downarrow$} & \textbf{MANIQA$\uparrow$} & \textbf{QAlign$\uparrow$} & \textbf{CLIPIQA$\uparrow$} & \textbf{NIQE$\downarrow$} & \textbf{MANIQA$\uparrow$} & \textbf{QAlign$\uparrow$} \\ \hline
Fast two-step &0.341 &\underline{3.903} &0.218 &3.913 &0.341 &\textbf{4.907} &0.227 &2.954 &0.324 &\textbf{3.835} &0.180 &3.393 \\
Universal IR &\textbf{0.470} &\textbf{3.548} &\textbf{0.310} &\textit{4.146} &\underline{0.407} &\textit{5.575} &\textit{0.282} &3.222 &\textbf{0.411} &\underline{3.843} &\textit{0.266} &\textit{3.621} \\
ZEBASELib-PT &0.336 &5.493 &0.211 &3.590 &0.319 &7.265 &0.213 &2.682 &0.299 &5.966 &0.202 &3.165 \\
OmniLens &\textit{0.383} &4.789 &\underline{0.289} &\underline{4.293} &\textit{0.392} &5.718 &\textbf{0.327} &\underline{3.899} &\textit{0.376} &\textit{4.311} &\underline{0.277} &\textbf{4.095} \\
\rowcolor{Gray}
\textbf{FoundCAC} &\underline{0.395} &\textit{4.374} &\textit{0.286} &\textbf{4.322} &\textbf{0.434} &\underline{5.296} &\underline{0.318} &\textbf{4.003} &\underline{0.378} &4.559 &\textbf{0.294} &\underline{3.727} \\ \hline \hline
\multirow{2}{*}{\textbf{Method}} & \multicolumn{4}{c|}{\textbf{Nano-Optics}} & \multicolumn{4}{c|}{\textbf{Canon 24mm f1/4}} & \multicolumn{4}{c}{\textbf{Sony 18135}} \\
 & \textbf{CLIPIQA$\uparrow$} & \textbf{NIQE$\downarrow$} & \textbf{MANIQA$\uparrow$} & \textbf{QAlign$\uparrow$} & \textbf{CLIPIQA$\uparrow$} & \textbf{NIQE$\downarrow$} & \textbf{MANIQA$\uparrow$} & \textbf{QAlign$\uparrow$} & \textbf{CLIPIQA$\uparrow$} & \textbf{NIQE$\downarrow$} & \textbf{MANIQA$\uparrow$} & \textbf{QAlign$\uparrow$} \\ \hline
Fast two-step &\textit{0.342} &\textbf{7.013} &0.231 &1.255 &\textit{0.469} &\textbf{3.292} &\textit{0.343} &\underline{4.318} &\textbf{0.560} &\textbf{2.826} &0.303 &\textit{4.380} \\
Universal IR &\textbf{0.389} &\underline{8.172} &0.263 &\textit{1.632} &0.433 &\underline{3.388} &0.332 &4.164 &0.489 &\underline{3.050} &\textit{0.317} &4.348 \\
ZEBASELib-PT &\underline{0.360} &9.522 &\textbf{0.299} &1.324 &\textbf{0.519} &4.182 &\underline{0.359} &\textit{4.301} &\underline{0.512} &4.264 &\textit{0.322} &\underline{4.387} \\
OmniLens &0.332 &\textit{8.735} &\underline{0.285} &\textbf{1.793} &\underline{0.492} &3.971 &\textbf{0.369} &\textbf{4.383} &\textit{0.499} &3.955 &\textbf{0.328} &\textbf{4.403} \\
\rowcolor{Gray}
\textbf{FoundCAC} &0.276 &9.689 &\textit{0.277} &\underline{1.701} &0.433 &\underline{3.388} &0.332 &4.164 &0.471 &\textit{3.945} &\underline{0.325} &\textbf{4.403} \\ \bottomrule[0.17em]
\end{tabular}
}
}
    \end{center}
\end{table*}

\section{More Experimental Results}
\label{sup:more results}
\subsection{Additional results for the evaluation on each lens of \textit{RealLens-Sim}}
\label{sup:simresults}
Table~\ref{tab:main_sup} reports the per‑lens quantitative results for each method as a complement to Table 2 of the main manuscript. 
Figures~\ref{fig:visual_sa}, \ref{fig:visual_mosmeta}, \ref{fig:visual_ma}, and \ref{fig:visual_smart} present the qualitative correction results of representative methods on selected \textit{RealLens‑Sim} lenses.
For lenses in MOS with more severe aberrations, where baseline methods still suffer from residual optical degradations, \ourframework~produces notably clearer results without introducing hallucinated artifacts, closely approximating the ground truth. 
For milder aberrations, such as those in MA cases and high‑end lenses, \ourframework~effectively refines image quality while strictly avoiding over‑sharpening, yielding highly natural results.
Overall, these additional results underscore the effective zero-shot generalizability of \ourframework.

\subsection{Additional analysis for supplemented specifications.}
As shown in Figure~\ref{fig:spe_od}, using Spatial‑Class, we visualize the optical degradation distributions of lens‑source samples generated by EAOD under the baseline specifications, with aspheric surface added, and with both aspheric surface (A.S.) and image plane perturbation (I.P.P.) added. The results are consistent with our motivation for introducing these factors, namely that both specifications lead the optimized lenses to exhibit new optical degradation patterns. Specifically, adding A.S. yields more samples with severe optical degradation because the increased number of optimization parameters makes optimization more difficult, while adding image‑plane perturbations yields more Spatial‑Class ``2'' and ``3'' samples because shifting the image plane position changes the in‑focus field.

\subsection{Visualization of the learned PSF representation}
To understand why LPR provides better guidance for aberration correction than the continuous PSF feature prediction baseline, we visualize their predicted spatial feature maps in Figure~\ref{fig:vis_lpr}. We compare two distinct optical degradation cases: S-1P-I (spatially non-uniform with moderate severity) and S-1P-II (spatially uniform with strong severity). 

Our proposed LPR produces distinctly different activation patterns that well align with the underlying physical degradation. Specifically, for S-1P-I, the LPR features exhibit a clearly spatially-varying activation pattern across different FoVs, whereas for S-1P-II, the activations are nearly global and consistent. This intuitively matches the spatial degradation distributions of the respective lenses, confirming that LPR effectively captures structured optical priors.

In contrast, directly predicting continuous PSF features struggles to clearly reflect these optical patterns. Without explicit structural constraints, the continuous features tend to entangle with image semantics (\textit{e.g.}, scene details and edges) rather than isolating pure aberrations. Consequently, their activation maps are less discriminative across different degradation cases. This visualization further validates the superiority of explicitly bounding the representation space with a discrete physical codebook.

\section{Additional Real-Capture, LPR, and RMS Analysis}
\label{sec:add_real_lpr_rms_analysis}

\begin{figure*}[!t]
  \centering
  \includegraphics[width=0.99\textwidth]{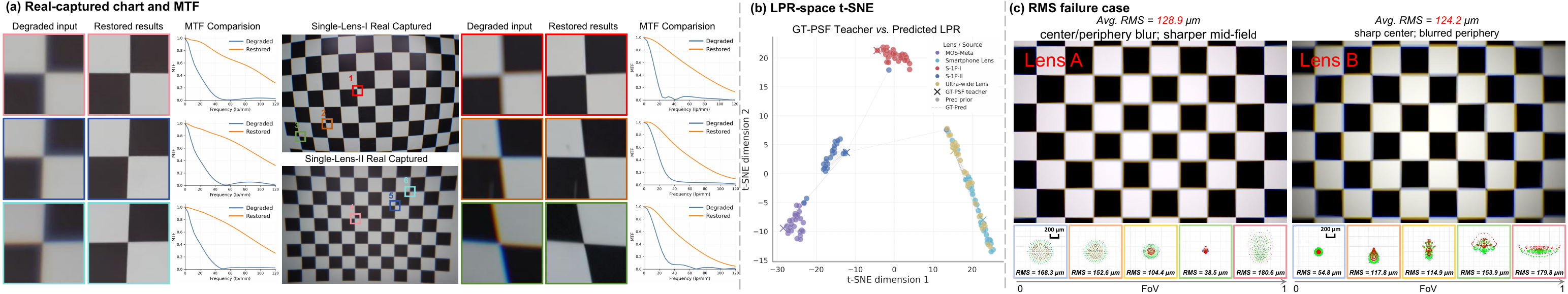}
  \caption{
  Additional real-capture and LPR analysis.
  (a) Representative checkerboard crops and image-domain edge-derived MTF curves. The MTF curves measure restored-image sharpness and may be affected by deconvolution/sharpening; thus they are complementary evidence rather than physical lens MTF or paired GT.
  (b) LPR-space visualization shows Pred--GT teacher alignment on unseen RealLens-Sim cases, while the prior coverage remains bounded by synthetic LensLib.
  (c) Similar average RMS can correspond to different spatial degradation patterns, showing that RMS alone is insufficient for LensLib sampling.
  }
  \label{fig:real_capture_lpr_analysis}
\end{figure*}

\begin{figure*}[!h]
  \centering
  \includegraphics[width=0.99\textwidth]{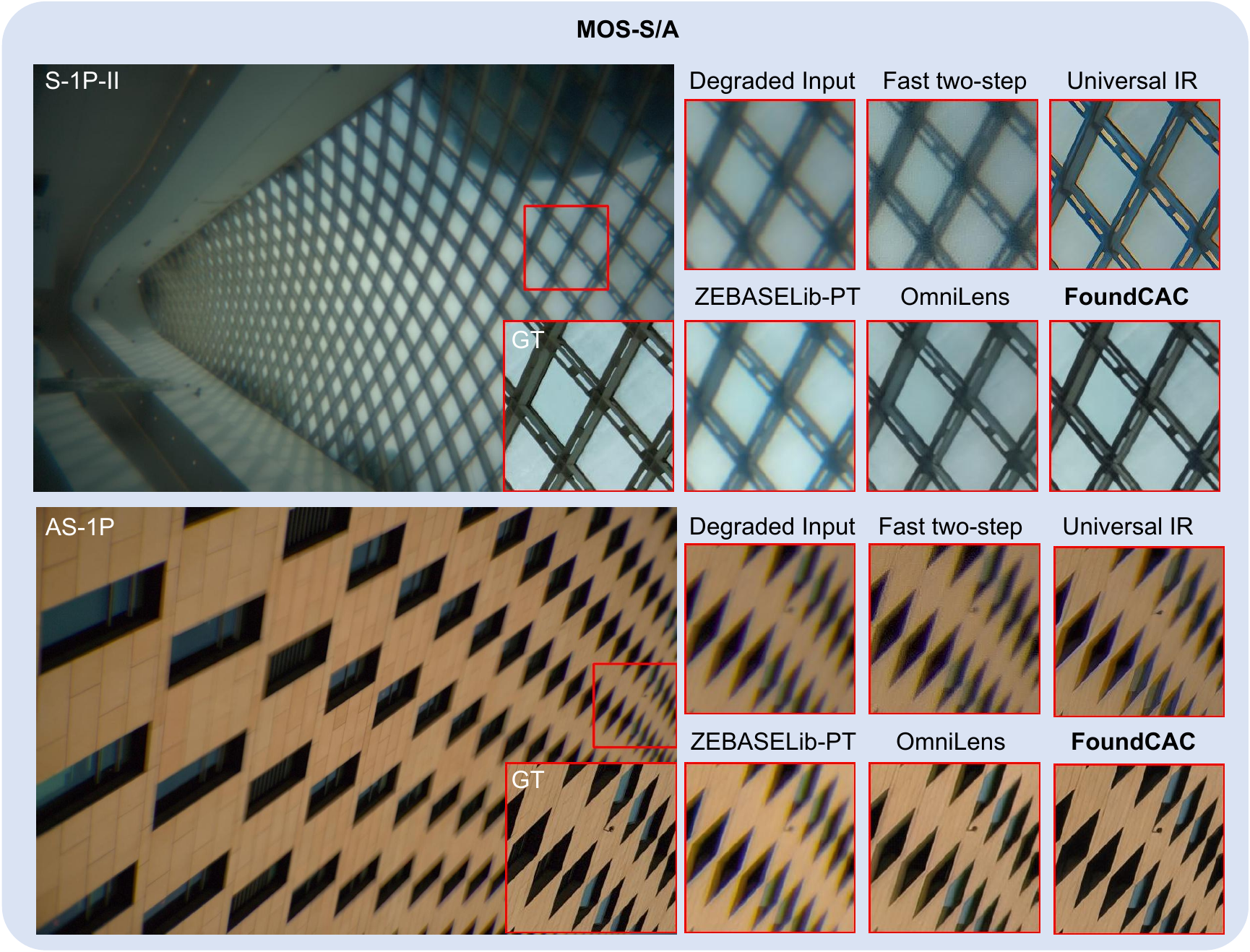}
  \caption{Visual comparison on MOS-S/A. S-1P-II and AS-1P are selected as the representative spherical and aspheric lenses for their distinct optical degradation patterns.}
  \label{fig:visual_sa}
\end{figure*}

\begin{figure*}[!h]
  \centering
  \includegraphics[width=0.99\textwidth]{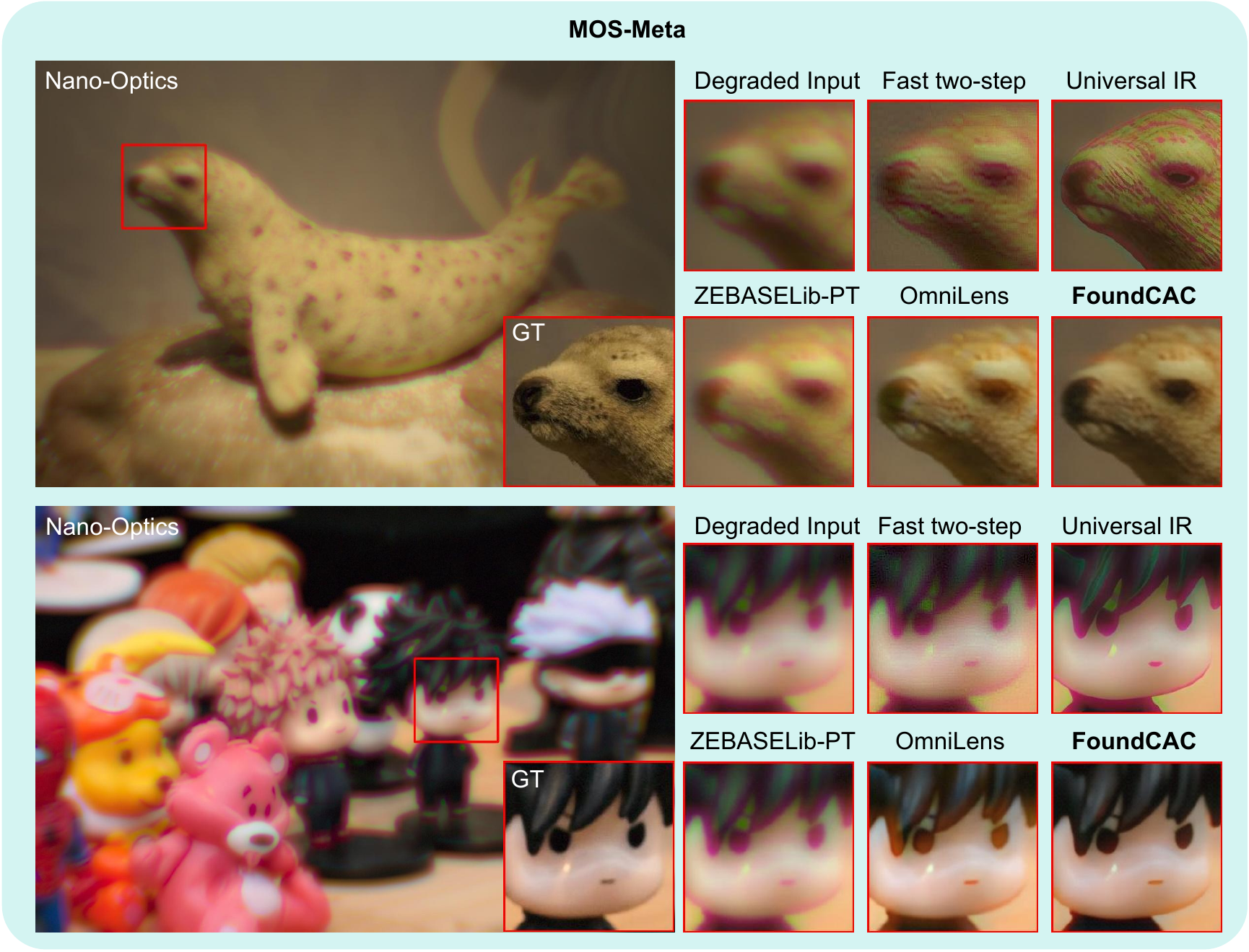}
  \caption{Visual comparison on MOS-Meta.}
  \label{fig:visual_mosmeta}
\end{figure*}

\begin{figure*}[!h]
  \centering
  \includegraphics[width=0.99\textwidth]{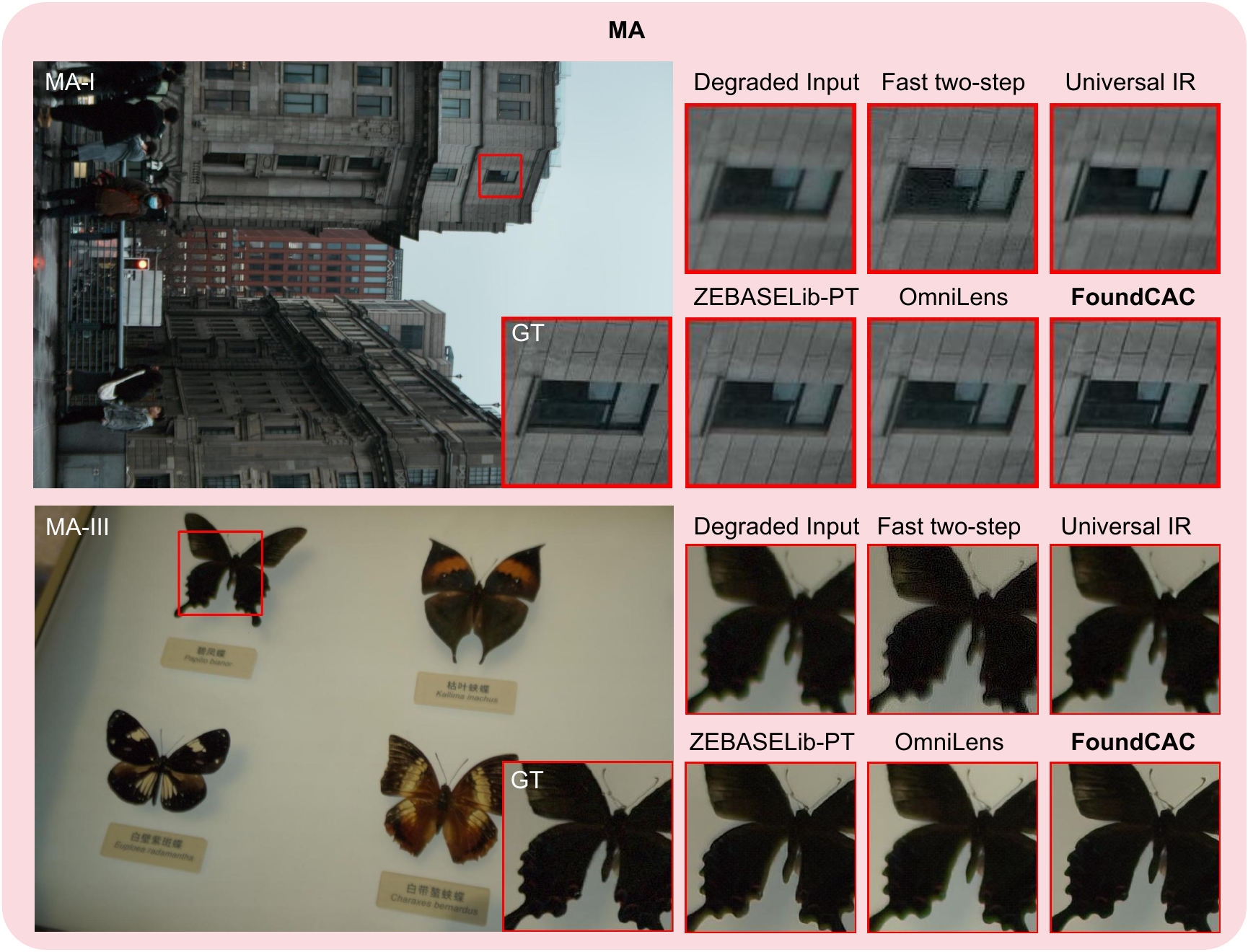}
  \caption{Visual comparison on MA. We present results under the minimum misalignment (MA-I) and maximum alignment (MA-III) settings.}
  \label{fig:visual_ma}
\end{figure*}

\begin{figure*}[!h]
  \centering
  \includegraphics[width=0.99\textwidth]{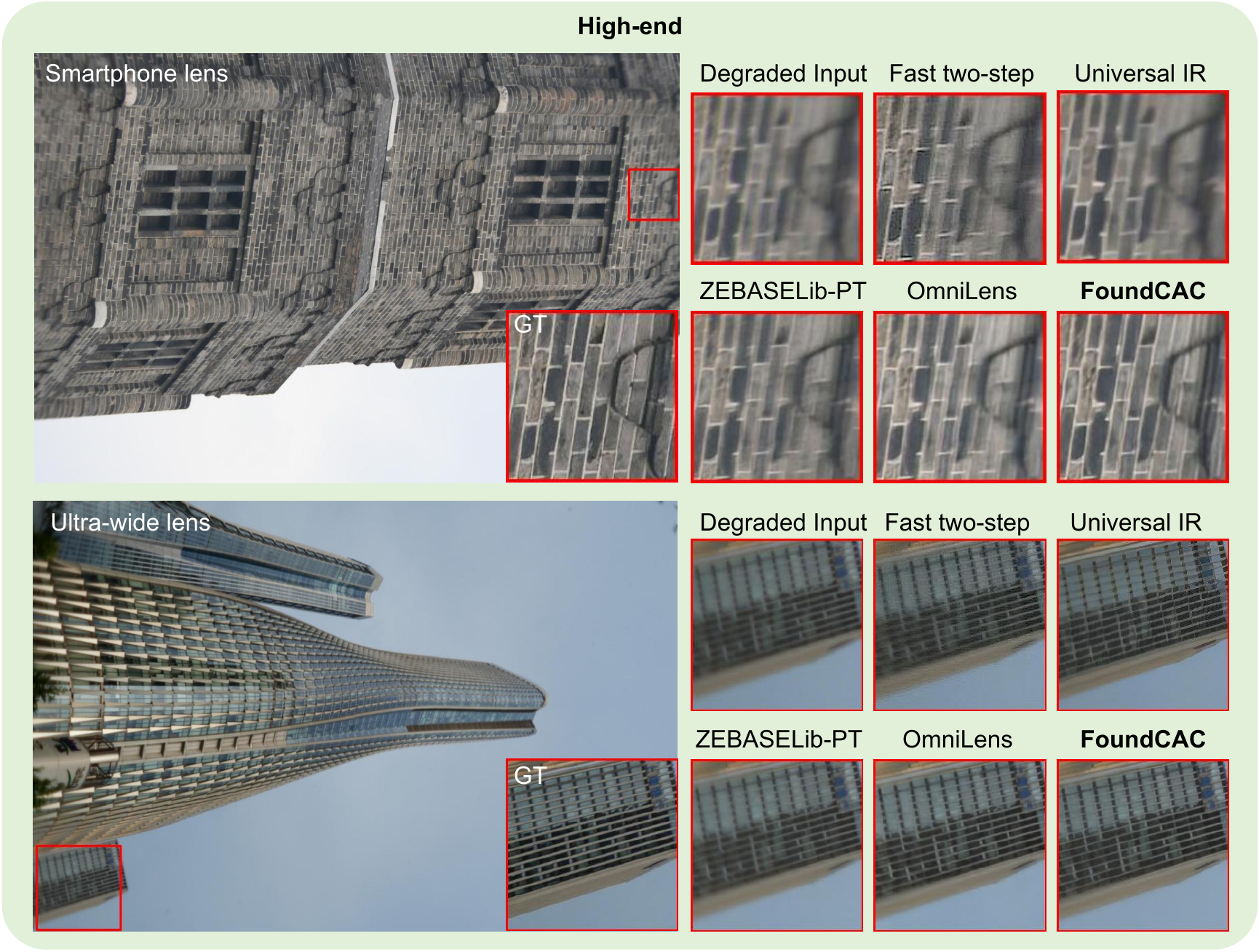}
  \caption{Visual comparison on High-end.}
  \label{fig:visual_smart}
\end{figure*}

We provide additional analyses to complement the main-paper evaluation and clarify the limitations discussed in the meta-review. 
Fig.~\ref{fig:real_capture_lpr_analysis}(a) shows representative real-captured checkerboard results with image-domain edge-derived MTF curves. These curves are computed from restored images and therefore reflect restored-image edge sharpness rather than the physical MTF of the lens. They should be interpreted as complementary chart-based evidence, not as paired full-reference real-capture validation.
Fig.~\ref{fig:real_capture_lpr_analysis}(b) visualizes the LPR space on unseen RealLens-Sim cases, where predicted priors tend to align with GT-PSF-derived teacher priors. This suggests that LPR learns a PSF-supervised latent structure, although its coverage is still bounded by the synthetic LensLib distribution.
Fig.~\ref{fig:real_capture_lpr_analysis}(c) illustrates an RMS failure case: lenses with similar average RMS can exhibit different spatial degradation patterns, supporting our use of severity/spatial-variation sampling rather than RMS-only selection.

\subsection{Numerical evaluation on \textit{RealLens-Snap}}
\label{sup:nriqa}

To provide a more comprehensive evaluation of representative blind aberration correction methods on real-captured data (\textit{RealLens-Snap}), we report non-reference image quality metrics, including CLIPIQA~\cite{wang2023exploring}, NIQE~\cite{mittal2012making}, MANIQA~\cite{yang2022maniqa}, and QAlign~\cite{wu2023q}, as summarized in Table~\ref{tab:nriqa}. 

Consistent with the visual results in Figure~4 of the manuscript, FoundCAC shows stable generalization across diverse real-world settings. On refractive optical systems (\textit{e.g.}, Single-Lens, CAYA, Canon, and Sony), its results rank within the Top~3 in most cases (16/20).

On the Nano-Optics lens, performance drops on CLIPIQA and NIQE, reflecting the domain gap between diffraction-dominated degradations and the ray-tracing-based priors used in our model. 
In contrast to data-driven methods that tend to produce conservative restorations, or generative approaches (\textit{e.g.}, Universal IR) that may alter image content and affect NR-IQA scores~\cite{zhang2024degradation}, our method constrains restoration within a refractive LPR space. Under diffraction-dominated conditions, this constraint may lead to suboptimal code selection and over-correction, which is penalized by certain metrics.

Despite this limitation, FoundCAC maintains competitive performance on QAlign, indicating that extending physically grounded priors to cover broader optical regimes remains an important direction for future work.

\subsection{Additional visual results on \textit{RealLens-Snap}}
\label{sup:real_visual}
\begin{figure*}[!h]
  \centering
  \includegraphics[width=0.99\textwidth]{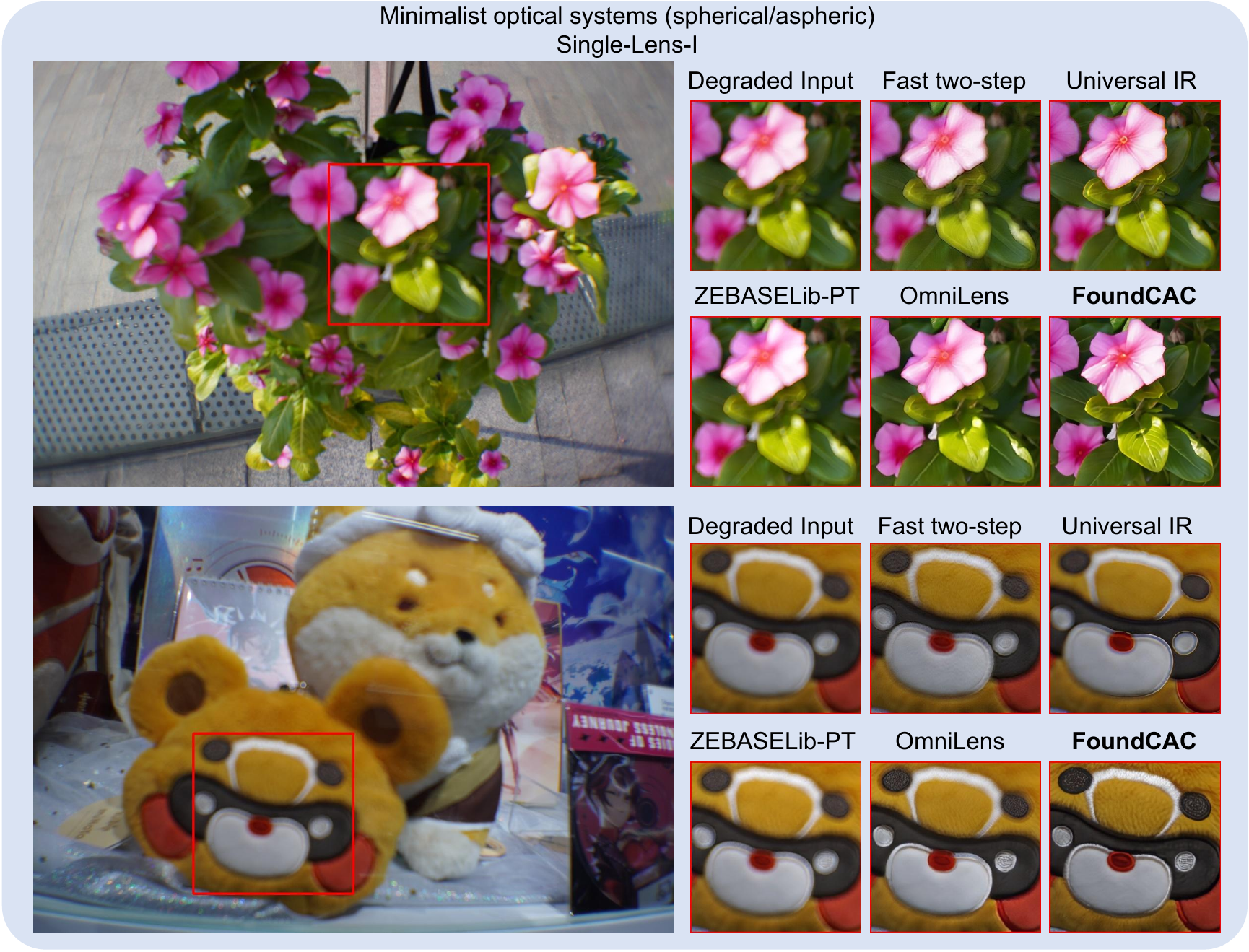}
  \caption{Visual comparison on Single-Lens-I.}
  \label{fig:visual_s1}
\end{figure*}

\begin{figure*}[!h]
  \centering
  \includegraphics[width=0.99\textwidth]{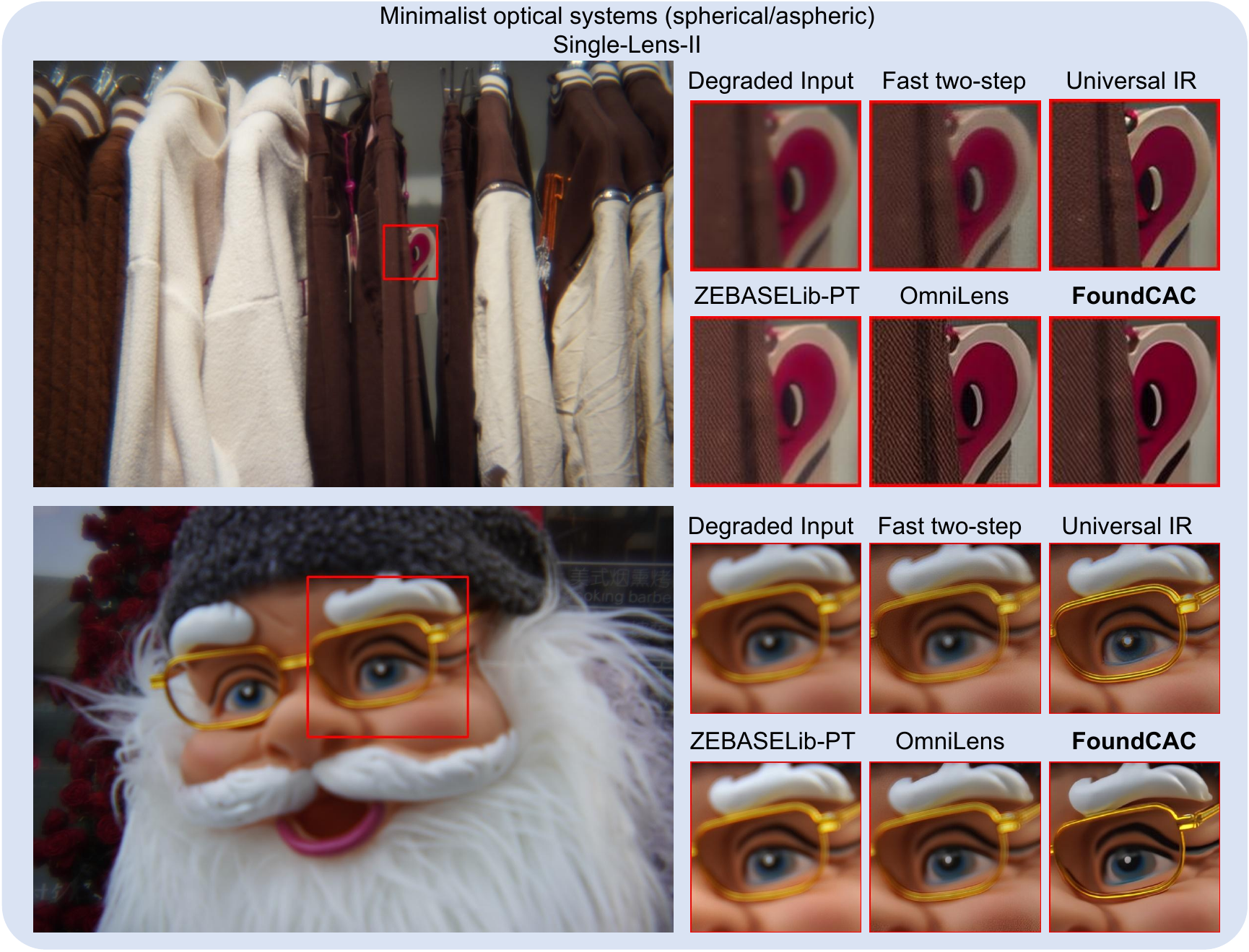}
  \caption{Visual comparison on Single-Lens-II.}
  \label{fig:visual_s3}
\end{figure*}

\begin{figure*}[!h]
  \centering
  \includegraphics[width=0.99\textwidth]{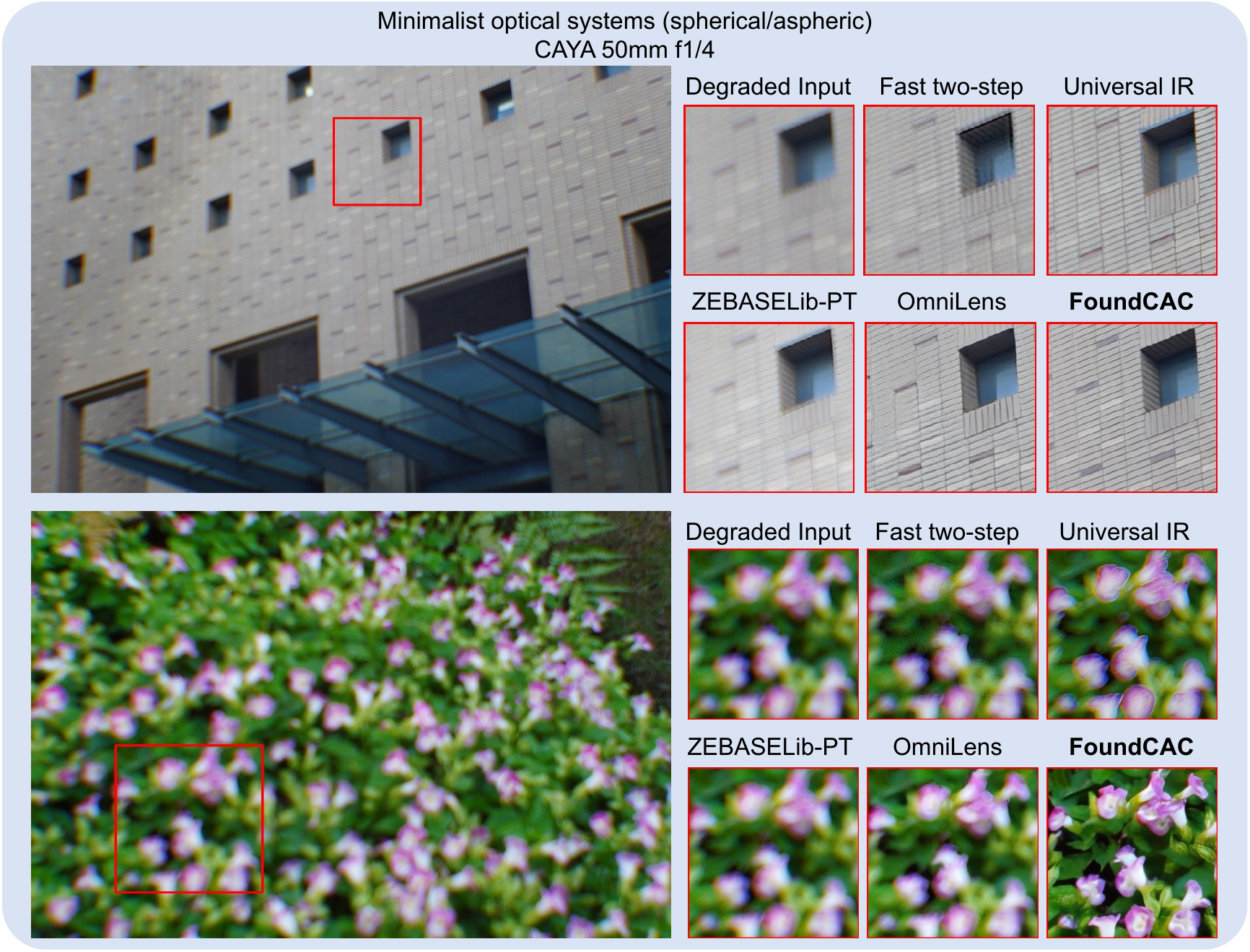}
  \caption{Visual comparison on $CAYA$ $50mm$ $f1/4$.}
  \label{fig:visual_caya}
\end{figure*}

\begin{figure*}[!h]
  \centering
  \includegraphics[width=0.99\textwidth]{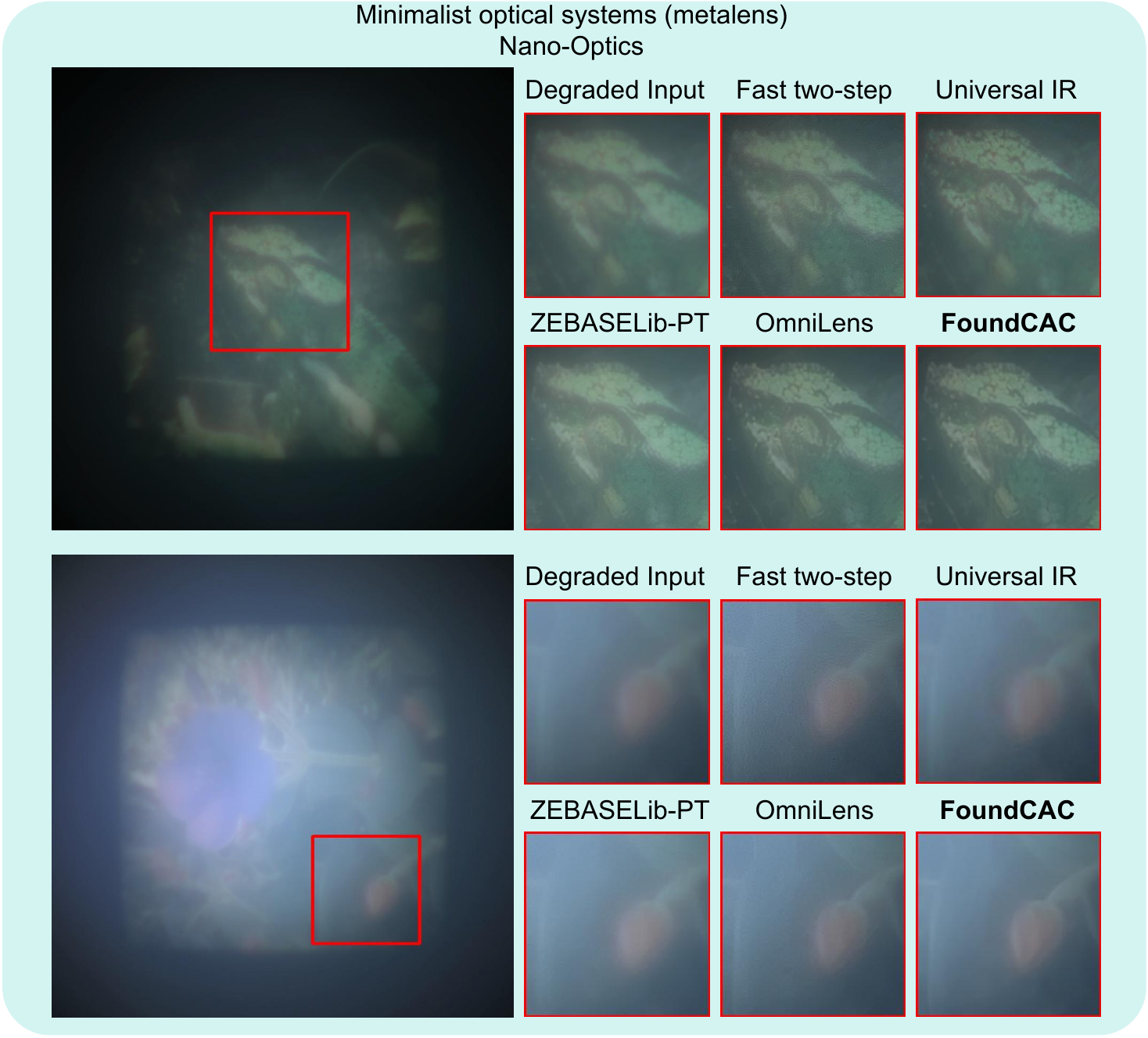}
  \caption{Visual comparison on Nano-Optics.}
  \label{fig:visual_nano}
\end{figure*}

\begin{figure*}[!h]
  \centering
  \includegraphics[width=0.99\textwidth]{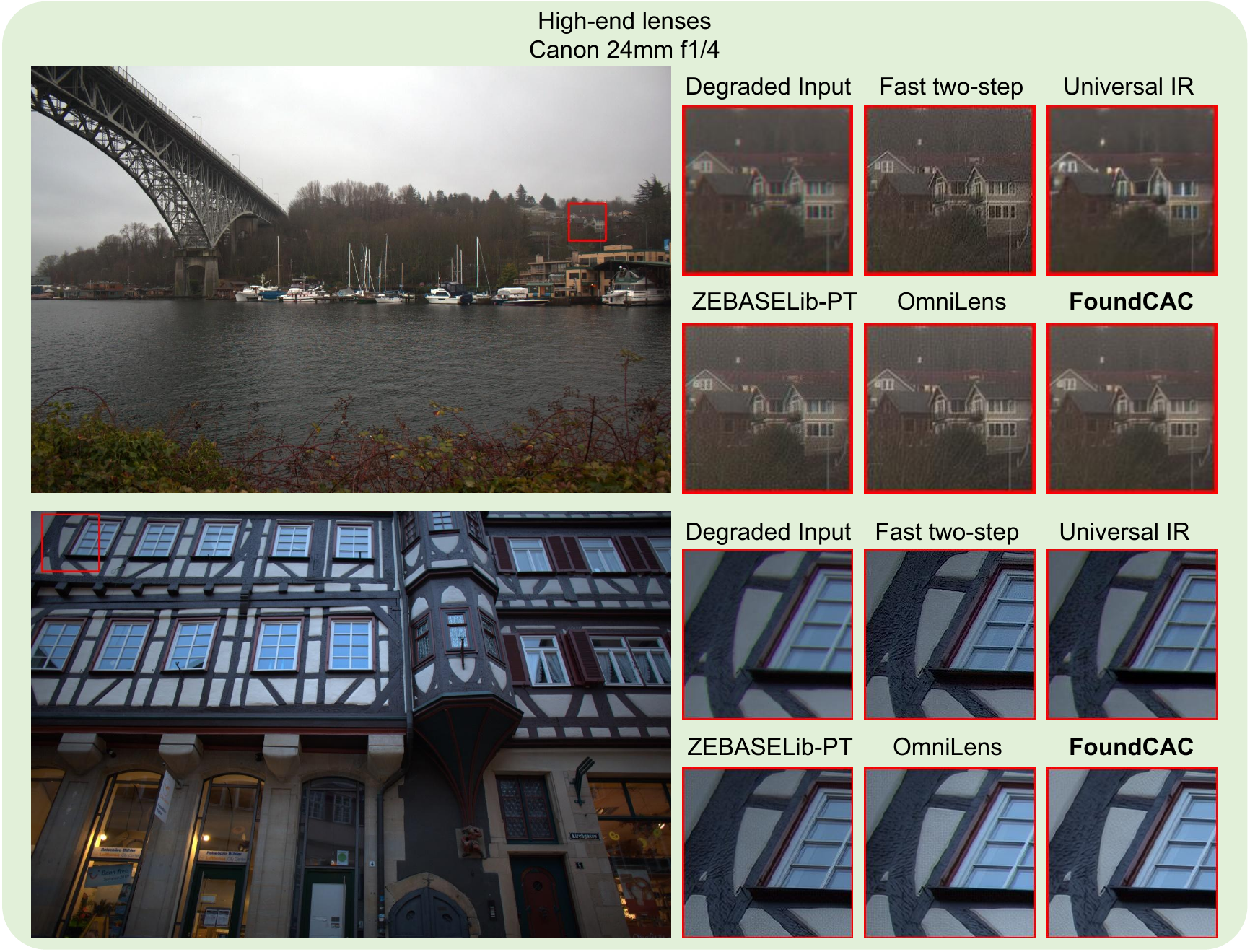}
  \caption{Visual comparison on $Canon$ $24mm$ $f1/4$.}
  \label{fig:visual_canon}
\end{figure*}

\begin{figure*}[!h]
  \centering
  \includegraphics[width=0.99\textwidth]{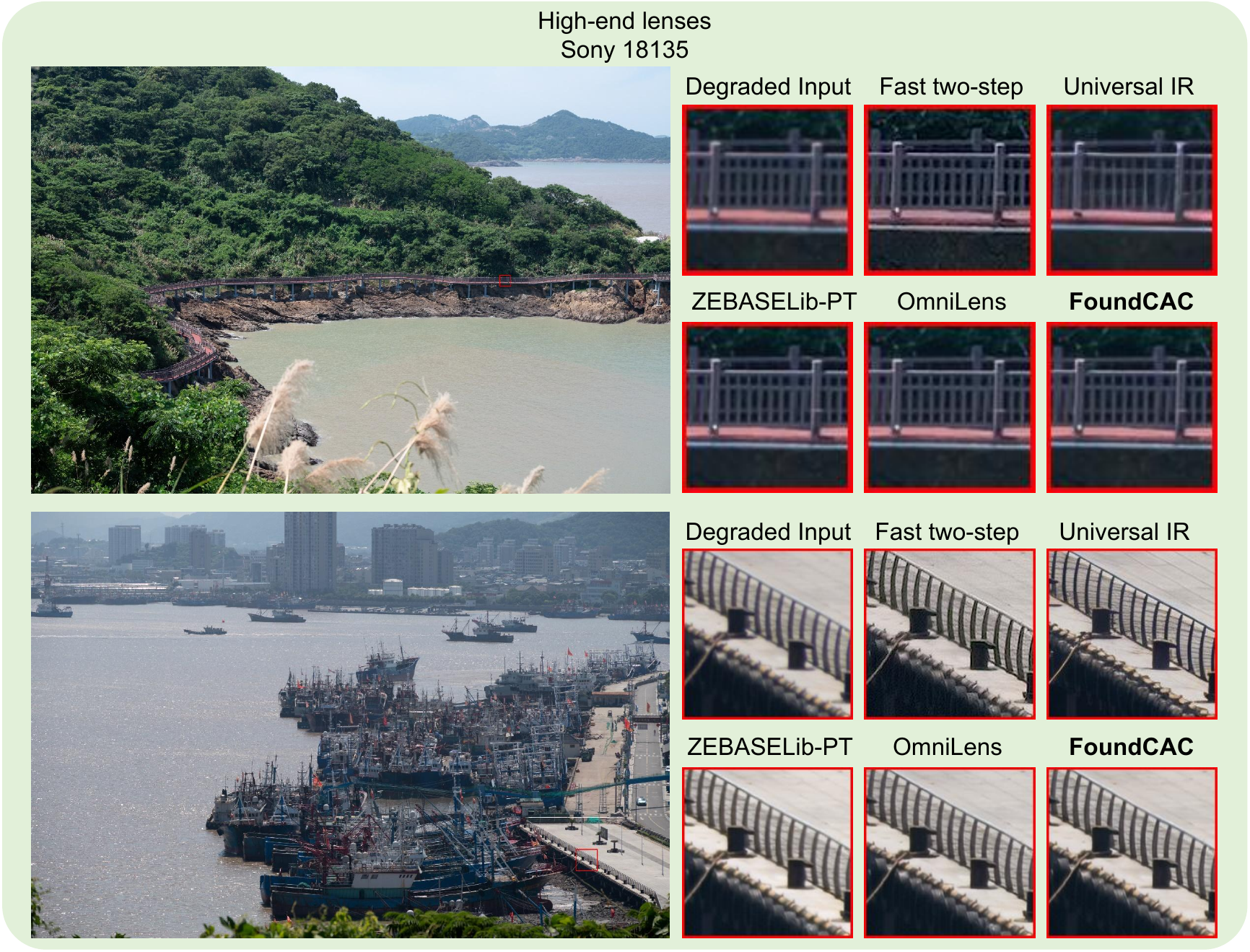}
  \caption{Visual comparison on $Sony$ $18135$.}
  \label{fig:visual_sony}
\end{figure*}

To further demonstrate the effectiveness of our method, we present additional comparison results between the proposed \ourframework~and competing approaches on real-world images captured with different physical lenses. 
Results on simple spherical and aspherical lenses are shown in Figures~\ref{fig:visual_s1},~\ref{fig:visual_s3}, and~\ref{fig:visual_caya}, those on metalens are shown in Figure~\ref{fig:visual_nano}, and those on high-end lenses are shown in Figures~\ref{fig:visual_canon} and~\ref{fig:visual_sony}.
The proposed \ourframework~consistently produces superior correction results across various lens types, further demonstrating its zero-shot capability in handling diverse real-world aberrations.
Specifically, for MOS, \ourframework~effectively mitigates severe aberrations and delivers high-fidelity correction results where competing methods perform unsatisfactorily.
For high-end DSLR lenses, it further alleviates residual degradations, including chromatic aberration, while strictly avoiding over-sharpening.

\clearpage

\end{document}